\documentclass[12pt]{article}
\pdfoutput=1

\usepackage{jheppub}
\usepackage{amssymb,amsfonts,amsmath,amsthm}
\usepackage{mathrsfs}
\usepackage{bbold}
\usepackage{tikz}
\usetikzlibrary{shapes.geometric, arrows}
\usepackage{comment}
\usepackage{tikz}

\makeatletter
\newcommand{\Vast}{\bBigg@{4.75}}
\makeatother

\newcommand{\be}{\begin{equation}}
\newcommand{\ee}{\end{equation}}
\newcommand{\bea}{\begin{eqnarray}}
\newcommand{\eea}{\end{eqnarray}}

\newcommand{\CB}{\mathcal{B}}

\newcommand{\CF}{\mathcal{F}}

\newcommand{\CL}{\mathcal{L}}
\newcommand{\CN}{\mathcal{N}}
\newcommand{\CM}{\mathcal{M}}

\newcommand{\CV}{\mathcal{V}}

\newcommand{\lr}{\left (}
\newcommand{\rr}{\right )}
\newcommand{\ls}{\left [}
\newcommand{\rs}{\right ]}
\newcommand{\lc}{\left \{}
\newcommand{\rc}{\right \}}

\newcommand\qt\tau

\newcommand{\p}{\partial}

\renewcommand{\tilde}[1]{\widetilde{#1}}

\makeatletter
\renewcommand{\@seccntformat}[1]{\csname the#1\endcsname.\,\,}
\makeatother
\allowdisplaybreaks

\let \savenumberline \numberline
\def \numberline#1{\savenumberline{#1.}}

\makeatletter
\def\@fpheader{\relax}
\makeatother

\def\bea{\begin{eqnarray}}
\def\eea{\end{eqnarray}}

\usetikzlibrary{shapes,arrows,chains}
\usetikzlibrary{decorations.markings}
\usetikzlibrary{decorations.pathmorphing}
\usetikzlibrary{shapes.multipart}
\tikzset{snake it/.style={decorate, decoration=snake}}

\title{\ \vspace{1.6cm} \\
\scalebox{0.97}{T-Duality in Nonrelativistic Open String Theory}}
\author{Jaume Gomis${}^a$, Ziqi Yan${}^{a,\,b}$, and Matthew Yu${}^a$}
\emailAdd{jgomis@perimeterinstitute.ca}
\emailAdd{ziqi.yan@su.se}
\emailAdd{myu@perimeterinstitute.ca}
\affiliation{
${}^a$Perimeter Institute for Theoretical Physics\\
31 Caroline St N, Waterloo, ON N2L 2Y5, Canada 
\medskip\\
${}^b$Nordita, KTH Royal Institute of Technology and Stockholm University\\
Roslagstullsbacken 23, SE-106 91 Stockholm, Sweden}
\abstract{Nonrelativistic open string theory  is defined  by a worldsheet theory  that produces   a Galilean invariant string spectrum   and is described at  low energies  
by a nonrelativistic Yang-Mills theory \cite{Gomis:2020fui}. We study T-duality transformations in the path integral  for the sigma model that describes nonrelativistic open string theory  coupled to an arbitrary closed string background, described by a string Newton-Cartan geometry, Kalb-Ramond, and dilaton field. We prove that    T-duality transformations   map nonrelativistic open string theory to relativistic and noncommutative open string theory in the discrete light cone quantization (DLCQ), a quantization scheme relevant for Matrix string theory. We also show how the    worldvolume  dynamics of nonrelativistic open string theory  described  by   the Dirac-Born-Infeld type action found in \cite{Gomis:2020fui} maps to the Dirac-Born-Infeld actions describing the worldvolume theories of the DLCQ of open string theory and noncommutative open string theory.   
}

\begin{document}

\maketitle
\vfill\eject
\section{Introduction and Conclusions}

Nonrelativistic string theory \cite{Gomis:2000bd} is described on the worldsheet by a two-dimensional relativistic quantum field theory (QFT).\footnote{The nonrelativistic spectrum was first obtained by taking a limit in \cite{Klebanov:2000pp} (see also \cite{Danielsson:2000gi}).} In a flat target spacetime, the worldsheet theory enjoys a nonrelativistic global symmetry,  dubbed as string Newton-Cartan symmetry (see e.g. \cite{Bergshoeff:2019pij}).  Realizing this symmetry requires introducing       one-form worldsheet fields, in addition to the worldsheet fields that play the role of target space coordinates. The nonrelativistic closed string spectrum is endowed with a Galilean invariant dispersion relation. 

The target space of dimension $d$ in nonrelativistic string theory is split into a two-dimensional longitudinal sector and a $(d-2)$-dimensional transverse sector, due to the presence of the one-form worldsheet fields. In   flat spacetime, the longitudinal sector is Lorentzian and the transverse sector is Euclidean. Nonrelativistic string theory is invariant under a Galilean-type boost symmetry that transforms the $d-2$ transverse directions into the two longitudinal directions, but not \emph{vice versa}.\footnote{The target space geometry of nonrelativistic string theory is the string Newton-Cartan geometry \cite{Bergshoeff:2018yvt} (see also \cite{Gomis:2005pg,Brugues:2004an,Andringa:2012uz,Brugues:2006yd,Bergshoeff:2019pij}). Note that this is different from nonrelativistic geometries derived from null reduction of relativistic string theory; see \cite{Harmark:2017rpg, Kluson:2018egd, Harmark:2018cdl, Harmark:2019upf}.} A crucial role is played by the topology of the spatial longitudinal coordinate: the closed  string spectrum is empty unless it is a circle  and the string can wind around the compact longitudinal coordinate \cite{Gomis:2000bd}.

Depending on whether a Dirichlet or Neumann boundary condition is imposed in the longitudinal spatial direction, there are two qualitatively different open string theories to consider (we take Neumann boundary conditions in the transverse directions). Imposing a Dirichlet boundary condition leads to nonrelativistic open string theory on  D$(d-2)$-branes, which has a nonrelativistic open string spectrum with a Galilean invariant dispersion relation \cite{Danielsson:2000mu,Gomis:2020fui}. 
Recently, in a companion paper \cite{Gomis:2020fui}, the worldsheet theory describing nonrelativistic open string theory (NROS) in an arbitrary open and closed string background has been put forward, and studied in detail. In particular, it was shown in \cite{Gomis:2020fui} that, in flat spacetime, the low energy dynamics of open strings ending on $n$ coinciding D$(d-2)$-branes is described by a Galilean invariant $U(n)$ Yang-Mills theory.\footnote{The theory on a single D-brane for $n=1$ is free and reduces to Galilean Electrodynamics \cite{Santos:2004pq, Bergshoeff:2015sic, Festuccia:2016caf, Chapman:2020vtn}.} Imposing a Neumann boundary condition instead leads to the well-studied noncommutative open string theory (NCOS) \cite{Gomis:2000bd, Danielsson:2000mu} on D$(d-1)$-branes,\footnote{In the original works on NCOS \cite{Seiberg:2000ms, Gopakumar:2000na}, the theory was defined as a zero slope, near critical electric field limit of relativistic open string theory. A finite worldsheet theory describing  NCOS was written down in \cite{Gomis:2000bd}.}
where a nonzero longitudinal $B$-field is required in order to have a nonempty open string spectrum. As a consequence of this, the longitudinal spacetime coordinates in NCOS do not commute with each other.

In this work we study T-duality of the path integral defining NROS. While T-duality in the transverse directions is in form the same as in relativistic open string theory, T-duality in the longitudinal spatial direction receives intriguing nonrelativistic twists. We will therefore focus on the longitudinal T-duality transformations in the following.

We first consider the path integral in flat spacetime in \S\ref{sec:flat} and uncover   the spacetime  interpretation  of the T-dual theories. We start with NROS on a D($d-2$)-brane, with a  Dirichlet boundary condition along the longitudinal spatial circle and Neumann boundary conditions in the remaining directions (and suitable boundary conditions for the one-form fields).  Performing a T-duality transformation along the longitudinal spatial circle in NROS leads to relativistic open string theory on a spacetime-filling D$(d-1)$-brane and with a compact lightlike circle.\footnote{Such a compact lightlike circle can be obtained by infinitely boosting a longitudinal spatial circle in the two-dimensional longitudinal sector \cite{Seiberg:1997ad}. The theory written in \cite{Gomis:2000bd} gives a finite worldsheet description of this limit.} The T-dual of NROS is thus the discrete light cone quantization (DLCQ) of relativistic open string theory (see \S\ref{sec:relnonrel}). 
This parallels the discussion of T-duality in nonrelativistic closed string theory \cite{Bergshoeff:2018yvt}.\footnote{Also see \cite{Kluson:2018vfd} for T-duality in nonrelativistic closed string theory in the Hamiltonian formalism.}
Quantizing string theory and M-theory in the DLCQ plays an important role in holographic correspondences such as the Matrix theory description of the DLCQ of M-theory \cite{Banks:1996vh} and of string theory \cite{Motl:1997th,Banks:1996my,Dijkgraaf:1997vv}.   Therefore, the worldsheet theory of NROS gives a first principles microscopic definition of open string theory in DLCQ.

We then proceed to consider a different T-duality transformation that relates NROS to NCOS. We note that, in longitudinal spatial T-duality, we have taken the geometry of the longitudinal sector in NROS to be a spacetime cylinder, wrapping around the compactified longitudinal spatial direction, and the D($d-2$)-brane is transverse to the longitudinal spatial circle. We can introduce a twist in the compactification of the longitudinal spatial coordinate by shifting one end of the longitudinal cylinder along the time direction before gluing back. Typically, this shift does not change the character of the T-duality transformation and still leads   to the DLCQ of relativistic open strings in the T-dual frame. However, when the shift equals the circumference of the longitudinal circle, 
performing a T-duality transformation on this shifted cylinder leads to NCOS on a spacetime-filling D($d-1$)-brane  and with a compact longitudinal lightlike circle. The T-dual theory is, in this sense, NCOS in the DLCQ description. The longitudinal $B$-field in the T-dual NCOS theory corresponds to a rescaling factor of the circle in NROS. See \S\ref{sec:nonrelnoncomm}.

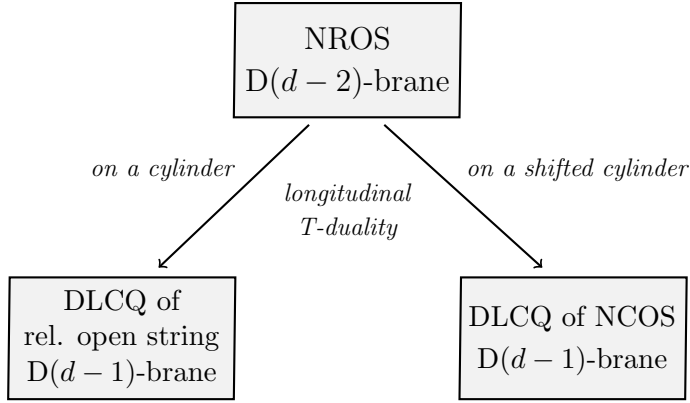
\begin{figure}[t!]\label{Section2Duality}
    \centering
\begin{tikzpicture}[thick]
    \def\Depth{3}
    \def\Height{2}
    \def\Width{2}
    \coordinate (O) at (0-5,0.1,0);
    \coordinate (A) at (.75-5,-.75,\Height);
     \coordinate (B) at (\Depth+.75-5,-.75,\Height);
    \coordinate (C) at (\Depth-5,0.1,0);
    \draw[black,fill=gray!10]   (-2.5+.5, 2.5+1+.25 ,0 ) -- (.75-2.5+.5,-.5+2.5+1,2) -- (3+0.75-2.5+.5,-.5+2.5+1,2) -- (3-2.5+.5,0+2.5+1+.25,0) -- cycle;
    \draw[black,fill=gray!10] (0+1, 0.1 ,0 ) -- (.75+1,-.5-.25,2) -- (3+0.75+1,-.5-.25,2) -- (3+1,0.1,0) -- cycle; 
    \draw[black,fill=gray!10] (O) -- (A) -- (B) -- (C) -- cycle;
    \draw[<-] (-2.25,1,2) -- (-.25,2.9,2);
    \draw[->]  (-.25+1,2.9,2) -- (2.85,1,2) ;
    \draw[right] (.95,-.5,.55) node{\begin{tabular}{c} \small DLCQ of NCOS \\ \small D$(d-1)$-brane\end{tabular}};
    \draw[right] (-1.5,3.70,1.9) node{\begin{tabular}{c} NROS \\ D$(d-2)$-brane\end{tabular}};
    \draw node at (-0.5,1.2) {\scalebox{0.8}{\emph{longitudinal}}};
    \draw node at (-0.5,0.75) {\scalebox{0.8}{\emph{T-duality}}};
    \draw[right] (-4.98,-.5,.5) node{\begin{tabular}{c} \small DLCQ of \\[-2.5pt]\small rel.  open string \\[-2.5pt] \small D$(d-1)$-brane\end{tabular}};
    \draw[right] (-4.1,1.75,.5) node{\begin{tabular}{c} \scalebox{0.8}{\textit{on a cylinder}} \end{tabular}};
    \draw[right] (0.9,1.75,.5) node{\begin{tabular}{c} \scalebox{0.8}{\textit{on a shifted cylinder}} \end{tabular}};
\end{tikzpicture}
\caption{Longitudinal T-duality  on a regular (shifted) cylinder relates NROS on a D($d-2$)-brane to the DLCQ of relativistic (noncommutative) open string theory on a spacetime-filling D($d-1$)-brane. Both of the dual theories on the lower left and right corners have a compact lightlike circle.}
\label{fig:triangle}
\end{figure}

The longitudinal T-duality transformation on a regular (shifted) cylinder that relates NROS to the DLCQ of relativistic (noncommutative) open string theory is summarized in Figure \ref{fig:triangle}.

 For completeness, we also consider in \S\ref{sec:noncommtorel} a longitudinal spatial T-duality transformation in the path integral of  nonrelativistic string theory with a spacetime-filling D$(d-1)$-brane, which leads to relativistic open string theory on a D($d-2$)-brane that is infinitely boosted along a spatial circle (see also \cite{Seiberg:2000ms}).  
 
Finally, in  \S\ref{sec:cs} we consider generalizations of all the T-duality transformations in general open and closed background fields, first on the worldsheet nonlinear sigma models and then on the associated spacetime Dirac-Born-Infeld type actions.

\section{Flat Spacetime} \label{sec:flat}

\subsection{Nonrelativistic Open String Theory} \label{sec:nos}

We first review nonrelativistic string theory in conformal gauge and with zero background fields, following closely \cite{Gomis:2020fui}. The free worldsheet action is
\be \label{eq:NROSfree}
	S_\text{NROS} = \frac{1}{4\pi\alpha'} \int_\Sigma d^2 \sigma \lr \p_\alpha X^{A'} \p^\alpha X^{A'} + \lambda \, \overline{\p} X + \overline{\lambda} \, \p \overline{X} \rr,
\ee 
where
\be
    X = X^0 + X^1,
        \qquad
    \overline{X} = X^0 - X^1,
\ee
and 
\be
	\p = \p_\sigma - i \, \p_\tau\,,
		\qquad
	\overline{\p} = \p_\sigma + i \, \p_\tau\,. 
\ee
Here, the Riemann surface $\Sigma$ is parametrized by the Euclidean coordinates $\sigma^\alpha = (\tau, \sigma)$\,. The target spacetime dimension is denoted by $d$\,, which is 26 for bosonic and 10 for supersymmetric nonrelativistic string theory. The worldsheet fields include the longitudinal coordinates $X^A$ with $A = 0\,, 1$\,, the transverse coordinates $X^{A'}$ with $A' = 2\,, \cdots, d-1$\,, and the one-form fields $\lambda$ and $\overline{\lambda}$. These one-form fields impose the (anti-)holomorphic constraints
$\overline{\p} X = \p \overline{X} = 0$\,.
The boundary $\p\Sigma$ of the worldsheet manifold $\Sigma$ is taken to be at $\sigma = 0$\,.  

In nonrelativistic open string theory, we impose a Dirichlet boundary condition in the longitudinal $X^1$-direction while taking all other directions to satisfy a Neumann boundary condition. Explicitly, the boundary conditions are
\begin{subequations} \label{eq:bdryconds}
\begin{align}
	\delta X^1 \big|_{\sigma = 0} & = 0\,, 
		&
	\quad \p_\sigma X^0 \, \big|_{\sigma = 0} = \p_\sigma X^{A'} \big|_{\sigma = 0} & = 0\,, \\[2pt]
	\lambda + \overline{\lambda} \, \big|_{\sigma = 0} & = 0\,,
		&
	\p_\sigma X^1 + i \, \p_\tau X^0 \, \big|_{\sigma = 0} & = 0\,.
\end{align}
\end{subequations}
These boundary conditions define nonrelativistic open string theory on a D($d$$-$2)-brane transverse to the longitudinal spatial $X^1$-direction. The symmetry preserved by the D-brane is the Bargmann symmetry.\footnote{There is also an additional dilatational symmetry acting on the longitudinal sector; see \cite{Gomis:2020fui}.}

This theory has an open string spectrum with a Galilean invariant dispersion relation. To have a nonempty spectrum, it is required that the longitudinal sector has a nontrivial topology; to be specific, we will consider the case where $X^1$ is compactified on a circle of radius $R$\,.\footnote{One may also consider open strings stretched between D-branes separated by a certain distance in the $X^1$ direction.} This implies the periodicity condition
\be
	X^1 (\sigma'+\pi) = X^1(\sigma') + 2\pi R \, w\,,
		\qquad
	w \in \mathbb{Z}\,,
\ee
where $w$ is the winding number. Here, $\sigma' = \tan^{-1} (\sigma/\tau)$\,.

\subsection{Longitudinal Spatial T-Duality} \label{sec:relnonrel}

In the following, we perform a T-duality transformation in nonrelativistic open string theory defined in \S\ref{sec:nos} along the $X^1$-direction, which is transverse to the D-brane. We start with introducing the parent action,
\begin{align} \label{eq:NROSparent}
	S_\text{parent} & = \frac{1}{4\pi\alpha'} \int_\Sigma d^2 \sigma \ls \p_\alpha X^{A'} \p^\alpha X^{A'} + \lambda \lr \overline{\p} X^0 + v_\sigma + i v_\tau \rr + \overline{\lambda} \lr \p X^0 - v_\sigma + i v_\tau  \rr \rs \notag \\[2pt]
	& \quad + \frac{i}{2\pi\alpha'} \int_{\Sigma} d^2 \sigma \, \epsilon^{\alpha\beta} \, \tilde{X} \, \p_\alpha v_\beta\,,
\end{align}
together with the boundary conditions
\begin{subequations} \label{eq:NROSparentbdryconds}
\begin{align}
	v_\tau \big|_{\sigma = 0} & = 0\,, 
		&
	\p_\sigma X^0 \, \big|_{\sigma = 0} = \p_\sigma X^{A'} \big|_{\sigma = 0} & = 0\,, \\[2pt]
	\lambda + \overline{\lambda} \, \big|_{\sigma = 0} & = 0\,,
		&
	v_\sigma + i \, \p_\tau X^0 \, \big|_{\sigma = 0} & = 0\,.
\end{align}
\end{subequations}
We have introduced the Lagrange multiplier $\tilde{X}$ that imposes $\epsilon^{\alpha\beta} \p_\alpha v_\beta = 0$ in the bulk; $\tilde{X}$ will be interpreted as a dual coordinate in the T-dual frame. Note that the theory is invariant under the global translation $\tilde{X} \rightarrow \tilde{X} + \xi$\,. To integrate out the   field $\tilde{X}$ in the path integral, we plug in \eqref{eq:NROSparent} the (local) solution to the bulk equation $\epsilon^{\alpha\beta} \p_\alpha v_\beta = 0$ with $v_\alpha = \p_\alpha X^1$.
This gives back the original action \eqref{eq:NROSfree} together with the boundary conditions \eqref{eq:bdryconds}. 

To proceed with the T-duality transformation, instead of integrating out $\tilde{X}$ in \eqref{eq:NROSparent}, we now integrate out the  field $v_\alpha$ in \eqref{eq:NROSparent}. The equations of motion from varying with respect to $v_\alpha$ are
\be \label{eq:dualitymap}
	\lambda + \overline{\lambda} = -  2 \, \p_\sigma \tilde{X}\,,
		\qquad
	\lambda - \overline{\lambda} = 2 \, i \, \p_\tau \tilde{X}\,,
\ee 
which define the duality map. Plugging \eqref{eq:dualitymap} into \eqref{eq:NROSparent} and \eqref{eq:NROSparentbdryconds}, we find the dual action
\be \label{eq:dualROS}
	\tilde{S}_\text{ROS} = \frac{1}{4\pi\alpha'} \int_\Sigma d^2 \sigma \lr \p_\alpha X^{A'} \p^\alpha X^{A'} - 2 \, \p_\alpha X^0 \, \p^\alpha \tilde{X} \rr, 
\ee
with the boundary conditions
\begin{align} \label{eq:dualbdryconds}
	\p_\sigma X^0 \, \big|_{\sigma = 0} = \p_\sigma \tilde{X} \big|_{\sigma = 0} = \p_\sigma X^{A'} \big|_{\sigma = 0} & = 0\,.
\end{align}
Note that the Dirichlet boundary condition $\delta X^1 \big|_{\sigma = 0} = 0$ in \eqref{eq:bdryconds} is dual to the Neumann boundary condition $\p_\sigma \tilde{X} \big|_{\sigma = 0} = 0$ in \eqref{eq:dualbdryconds}.  

Define $\tilde{X}^\mu$ with $\mu = (A, A')$ and
\be \label{eq:dualcoordinate}
	\tilde{X}^0 \equiv \frac{1}{\sqrt{2}} \bigl( X^0 + \tilde{X} \bigr)\,, 
		\qquad
	\tilde{X}^1 \equiv \frac{1}{\sqrt{2}} \bigl( X^0 - \tilde{X} \bigr)\,.
		\qquad
	\tilde{X}^{A'} = X^{A'}.
\ee
In terms of $\tilde{X}^\mu$, the dual action \eqref{eq:dualROS} becomes
\be \label{eq:tildeSROS}
	\tilde{S}_\text{ROS} = \frac{1}{4\pi\alpha'} \int_\Sigma d^2 \sigma \, \p_\alpha \tilde{X}^\mu \, \p^\alpha \tilde{X}_\mu\,,
\ee
together with the Neumann boundary condition $\p_\sigma \tilde{X}^\mu \big|_{\sigma = 0} = 0$\,. This is relativistic open string theory on a spacetime-filling D$(d-1)$-brane 
in Minkowski space, in which the spacetime coordinate $\tilde{X}$ is a compact lightlike direction. This yields the DLCQ of relativistic open string theory. See more in \S\ref{sec:osvo}.

It is also interesting to perform the inverse T-duality transformation that takes \eqref{eq:tildeSROS} back to \eqref{eq:NROSfree}. We start with introducing the parent action
\begin{align} \label{eq:parentaction0}
    \tilde{S}_\text{parent} & = \frac{1}{4\pi\alpha'} \int_\Sigma d^2 \sigma \, \Bigl( \p_\alpha X^{A'} \, \p^\alpha X^{A'} - 2 \, u_\alpha \, \p^\alpha X^0 \, \Bigr) \notag \\[2pt]
    & \quad + \frac{i}{2\pi\alpha'} \int_\Sigma d^2 \sigma \, \epsilon^{\alpha\beta} \, Y \p_\alpha u_\beta + \frac{i}{2\pi\alpha'} \int_{\p\Sigma} d\tau \, Y \, u_\tau \,,
\end{align}
with the boundary conditions
\be \label{eq:dualbdryconds1}
	\p_\sigma X^0 \big|_{\sigma = 0} = \p_\sigma X^{A'} \big|_{\sigma = 0} = 0\,, 
		\qquad
	u_\sigma \big|_{\sigma = 0} = 0\,, 
		\qquad
	\delta Y \big|_{\sigma = 0} = 0\,.
\ee
Here, $Y$ is a Lagrange multiplier that imposes $\epsilon^{\alpha\beta} \p_\alpha u_\beta = 0$\,. The boundary term $Y \, u_\tau$ is introduced so that the action is invariant under the global translation $Y \rightarrow Y + \xi$\,, reflecting the fact that it is immaterial where along the $Y$ direction the dual brane is exactly located \cite{Alvarez:1996up}. Integrating out the Lagrange multiplier $Y$ takes the action \eqref{eq:parentaction0} back to \eqref{eq:dualROS}. To pass to the T-dual theory, we rewrite \eqref{eq:parentaction0} as
\begin{align} \label{eq:freerelaction1}
    \tilde{S}_\text{parent} & = \frac{1}{4\pi\alpha'} \int_\Sigma d^2 \sigma \, \p_\alpha X^{A'} \, \p^\alpha X^{A'} 
    - \frac{1}{2\pi\alpha'} \int_\Sigma d^2 \sigma \, u_\alpha \bigl( \p^\alpha X^0 - i \, \epsilon^{\alpha\beta} \p_\beta Y \, \bigr)\,.
\end{align}
The auxiliary field $u_\alpha$ serves as a Lagrange multiplier which imposes 
\be \label{eq:constraints0}
	\p^\alpha X^0 - i \, \epsilon^{\alpha\beta} \p_\beta Y = 0\,.
\ee
Identifying 
\be \label{eq:lambdau}
	\lambda = - u_\sigma + i \, u_\tau\,, 
		\qquad
	\overline{\lambda} = - u_\sigma - i \, u_\tau\,,
\ee
and $Y = X^1$, we find that \eqref{eq:freerelaction1} takes the form of \eqref{eq:NROSfree} and the boundary conditions in \eqref{eq:dualbdryconds1} reduce to the ones in \eqref{eq:bdryconds}.

\subsection{Open String Vertex Operators and  T-Duality} \label{sec:osvo}

The most general open string vertex operator in nonrelativistic open string theory has been constructed in \cite{Gomis:2020fui}, with
\be \label{eq:osvo}
	\CV_\text{NROS} = \int_{\p\Sigma} d\tau \, \bigl[ N \lambda + i \bigl( A_0 \, \p_\tau X^0 + A_{A'} \, \p_\tau X^{A'} \bigr) \bigr]\,. 
\ee
For zero winding states, the worldsheet couplings $N$, $A_0$ and $A_{A'}$ are functions of $X^0$ and $X^{A'}$\,. Here, $A_0$ and $A_{A'}$ are components of the $U(1)$ gauge field living on the D($d-2$)-brane; $N$ is the Nambu-Goldstone boson that perturbs the geometry of the brane, and it arises due to the fact that the D($d-2$)-brane spontaneously breaks the translational symmetry in $X^1$. They have spacetime interpretation as background open string fields on the D($d$$-$2)-brane. This vertex operator is invariant under that Bargmann symmetry. The couplings are independent of $\lambda$ and the vertex operator is invariant under the $U(1)$ gauge transformation
\be
	\delta_\epsilon A_0 = \p_0 \epsilon\,,
		\qquad
	\delta_\epsilon A_{A'} = \p_{A'} \epsilon\,.
\ee
In nonrelativistic open string theory, open string states with a fixed winding number $w$ are described by a vertex operator which includes the following factor \cite{Gomis:2000bd}: 
\be \label{eq:windingvo}
	V^w = \exp \ls - \frac{i w R}{\alpha'} \int^z dz' \, \lambda(z') \rs, 
		\qquad
	z = \sigma + i \, \tau.
\ee
Then, the couplings $N$, $A_0$ and $A_{A'}$ are $\lambda$-dependent and can be ``Fourier expanded":
\begin{subequations}
\begin{align}
    N (X^0, X^{A'}, \lambda) & = \sum_{w} N^w (X^0, X^{A'}) \, V^w, \\[2pt]
    A_0 (X^0, X^{A'}, \lambda) & = \sum_{w} A_0^w (X^0, X^{A'}) \, V^w, \\[2pt]
    A_{A'} (X^0, X^{A'}, \lambda) & = \sum_w A_{A'}^w (X^0, X^{A'}) \, V^w.
\end{align}
\end{subequations} 
Therefore, for winding states, the vertex operator is invariant under the $U(1)$ gauge transformation that acts on the ``Fourier modes" as
\be \label{eq:gaugeNw}
	\delta_\epsilon N^w = i \, \frac{w R}{\alpha'} \, \epsilon^w\,, 
		\qquad
	\delta_\epsilon A^w_0 = \p_0 \, \epsilon^w\,,
		\qquad
	\delta_\epsilon A^w_{A'} = \p_{\!A'} \epsilon^w\,.
\ee
The fact that there is a nontrivial $U(1)$ transformation acting on $N^w$ will gain a simple interpretation in the T-dual frame.  

In the parent action \eqref{eq:NROSparent}, the open string vertex operator \eqref{eq:osvo} remains the same. Integrating out $v_\alpha$ in the path integral amounts to applying the duality map \eqref{eq:dualitymap} together with the dual boundary condition \eqref{eq:dualbdryconds} to the vertex operator \eqref{eq:osvo}. As a result, the T-dual of the open string vertex operator \eqref{eq:osvo} is
\be
	\tilde{\CV}_\text{ROS} = i \int_{\p\Sigma} d\tau \, \bigl( N \, \p_\tau \tilde{X} + A_0 \, \p_\tau X^0 + A_{A'} \p_\tau X^{A'} \bigr)\,. 
\ee 
In terms of the dual coordinates defined in \eqref{eq:dualcoordinate}, we find
\be
	\tilde{\CV}_\text{ROS} = i \int_{\p\Sigma} d\tau \, \tilde{A}_\mu \, \p_ \tau \tilde X^\mu,
\ee
where $\tilde{X}^\mu$ is defined in \eqref{eq:dualcoordinate} and
\be \label{eq:gtA01}
	\tilde{A}_0 = \frac{1}{\sqrt{2}} \bigl( N + A_0 \bigr)\,,
		\qquad
	\tilde{A}_1 = \frac{1}{\sqrt{2}} \bigl( - N + A_0 \bigr)\,.
\ee
This is the standard open string vertex operator in relativistic open string theory. Here, $\tilde{A}_\mu$ is the $U(1)$ gauge field on the spacetime-filling brane, which transform under the gauge symmetry as
\be \label{eq:gaugeros}
	\delta_{\tilde{\epsilon}} \, \tilde{A}_\mu = \frac{\p \, \tilde{\epsilon}}{\p \tilde{X}^\mu}\,.
\ee  

Next, we consider the T-dual of the winding operator \eqref{eq:windingvo}. Using the duality map \eqref{eq:dualitymap}, we find the dual of \eqref{eq:windingvo} as $\exp \bigl( i \, \tilde{p} \, \tilde{X} \bigr)$, where
\be \label{eq:KKwinding}
	\tilde{p} = \frac{\tilde{n}}{\tilde{R}}\,, 
		\qquad
	\tilde{n} = w\,, 
		\qquad
	\tilde{R} = \frac{\alpha'}{R}\,.
\ee
In order for $\exp \bigl( i \, \tilde{p} \, \tilde{X} \bigr)$ to be a physical operator, it is required that
\be
	\tilde{X} (\sigma' + \pi) = \tilde{X} (\sigma') + 2 \pi \tilde{R} \, \tilde{w}\,, 
		\qquad
	\tilde{w} \in \mathbb{Z}\,.
\ee
Therefore, the dual coordinate $\tilde{X}$ is compactified on a lightlike circle of radius $\tilde{R}$\,, and it can carry a Kaluza-Klein excitation number $\tilde{n}$\,. The equation $\tilde{n} = w$ in \eqref{eq:KKwinding} gives the usual mapping between the Kaluza-Klein excitation and winding number. This observation implies that the dual theory is the discrete light cone quantization (DLCQ) of relativistic open string theory.

We now return to the gauge transformation of $N^w$ in \eqref{eq:gaugeNw}. In the T-dual frame, we have $N = \bigl( \tilde{A}_0 - \tilde{A}_1 \bigr) / \sqrt{2}$ from \eqref{eq:gtA01}, and the $U(1)$ gauge transformation of $N$ follows from \eqref{eq:gaugeros}, with 
\be \label{eq:gaugeNdual}
	\delta_{\tilde{\epsilon}} \, N = \frac{\p \, \tilde{\epsilon}}{\p \tilde{X}} \,,
\ee
Taking the Fourier expansion of $N$ with respect to the dual coordinate $\tilde{X}$\,, we define
\begin{subequations}
\begin{align}
	N (X^0, X^{A'}, \tilde{X}) & = \sum_{\tilde{n}} N^{\tilde{n}} (X^0\,, X^{A'}) \, \exp \bigl( i \, \tilde{p} \, \tilde{X} \bigr)\,, \\[2pt]
	\tilde{\epsilon} (X^0, X^{A'}, \tilde{X}) & = \sum_{\tilde{n}} \epsilon^{\tilde{n}} (X^0, X^{A'}) \, \exp \bigl( i \, \tilde{p} \, \tilde{X}\bigr)\,.
\end{align}
\end{subequations}
Then, \eqref{eq:gaugeNdual} implies that
$\delta_\epsilon N^{\tilde{n}} = i \, \tilde{p} \, \epsilon^{\tilde{n}}$\,,
which matches with the transformation in \eqref{eq:gaugeNw} after identifying $\epsilon^{\tilde{n}}$ with $\epsilon^w$ and applying \eqref{eq:KKwinding}. In the dual frame, the Nambu-Goldstone boson $N$ becomes a component of the $U(1)$ gauge field.

\subsection{Longitudinal T-Duality on a Shifted Cylinder} \label{sec:nonrelnoncomm}

In this subsection, we consider a different T-duality transformation in nonrelativistic open string theory, for which the dual theory gives rise to noncommutative open string theory (NCOS) in flat spacetime. 

We start with the same worldsheet action \eqref{eq:NROSfree} that describes nonrelativistic open strings,
\be \label{eq:NROSfree2}
	S_\text{NROS} = \frac{1}{4\pi\alpha'} \int_\Sigma d^2 \sigma \lr \p_\alpha X^{A'} \p^\alpha X^{A'} \! + \lambda \, \overline{\p} X + \overline{\lambda} \, \p \overline{X} \rr,
\ee 
and rewrite the associated boundary conditions in \eqref{eq:bdryconds} equivalently as
\begin{subequations} \label{eq:bdry3}
\begin{align}
    \p_\sigma X^{A'} \big|_{\sigma = 0} & = 0\,,
        &%
    \lambda + \overline{\lambda} \, \big|_{\sigma = 0} & = 0\,, \\[2pt]
    \overline{\p} X \big|_{\sigma = 0} = \p \overline{X} \big|_{\sigma = 0} & = 0\,, 
        & %
    \p X + \overline{\p X} \, \big|_{\sigma = 0} & = 0\,.
\end{align}
\end{subequations}
Note that the boundary conditions $\overline{\p} X + \p \overline{X} \big|_{\sigma = 0}=0$ and $\p X + \overline{\p X} \, \big|_{\sigma = 0} = 0$ in \eqref{eq:bdry3} are equivalent to the Dirichlet boundary condition $\p_\tau X^1 \big|_{\sigma = 0} = 0$ and the Neumann boundary condition $\p_\sigma X^0 \big|_{\sigma = 0} = 0$\,. These boundary conditions imply that there is a D($d-2$)-brane transverse to the longitudinal spatial $X^1$-direction. 

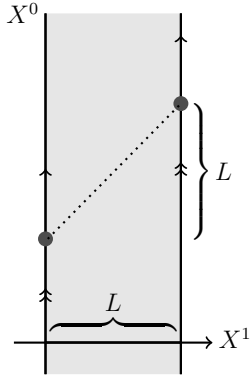
\begin{figure}[t!]
 \centering
    \begin{tikzpicture}[thick, scale=0.6]
     \tikzstyle{vertex}=[circle,fill=black!70, minimum size=2pt,inner sep=2pt]
     \draw[fill=black!10, opaque = .2] (-2,5)--(-2,-2.3)--(1,-2.3) -- (1,5);
     \draw[fill=black!10, opaque = .2] (-2,-3)--(-2,-2.3) -- (1,-2.3)--(1,-3);
    \draw[decoration={markings, mark=at position 0.45 with {\arrow{<}},mark=at position 0.8 with {\arrow{<<}}}, postaction={decorate},thick] (-2,5) -- (-2,-3);
    \draw[decoration={markings, mark=at position 0.08 with {\arrow{<}},mark=at position 0.45 with {\arrow{<<}}}, postaction={decorate},thick] (1,5) -- (1,-3);
    \node[vertex]at (-2, -1+1) {};
    \node[vertex]at (1, 2+1) {};
    \draw[black,dotted] (-1.95 , -1+1) --( 1 , 2+1);
    \draw[->] (-2.7,-2.3) -- (1.7,-2.3);
    \draw
    (2.2,-2.2) node {\scalebox{0.8}{$X^1$}};
    \draw (-2.5,5) node {\scalebox{0.8}{$X^0$}};
    \draw (1.45,1.5) node {\scalebox{0.8}{$\Vast\}$}};
    \draw (1.95,1.5) node {\scalebox{0.8}{$L$}};
    \draw (-0.5,-1.9) node {\scalebox{0.8}{$\overbrace{\hspace{2.1cm}}$}};
    \draw (-0.5,-1.4) node {\scalebox{0.8}{$L$}};
    \end{tikzpicture}
    \caption{The geometry of the longitudinal spacetime as a cylinder with a shift on one end.  The two end points of the dashed line are identified.}
    \label{fig:sc}
\end{figure}

In the previous subsections, we assumed that $X^1$ is compactified on a longitudinal spatial circle of radius $R$\,, and the longitudinal sector of the target space has a cylindrical geometry. Now, we introduce a shift on one end of this cylinder by $L = 2 \pi R$ in the $X^0$-direction as shown in Fig. \ref{fig:sc}, before identifying it with the other end. The shifted cylinder is defined to be glued using the following equivalence relation:
\be \label{eq:ec}
    \bigl( X^0, X^1 \bigr) : \,\, \bigl( X^0, 0 \bigr) \sim \bigl( X^0 \! + L\,, 2\pi R \bigr)\,,
\ee
with $X^0 \in \mathbb{R}$ and $X^1 \in [0,2\pi R]$\,.
Under this shift in the $X^0$ direction, the longitudinal spatial circle is tilted to be the dashed line in Fig. \ref{fig:sc}, with its two ends identified. It is important to note that $L = 2 \pi R$ (or $L = - 2 \pi R$) is a very special choice: only in this case does the line joining between two identified points on the cylinder (such as the dashed line in Fig. \ref{fig:sc}) lie along a lightlike direction.

To perform a T-duality transformation on this shifted cylinder, it is useful to take the following change of variables:
\be \label{eq:cvlp}
    \lambda' = E^{-1} \lambda\,,
        \qquad
    Y = 2 E X^1,
        \qquad
    \overline{Y} = X^0 - X^1,
\ee
where we introduced a nonzero constant parameter $E$\,. Under this change of variables, we find that the tilted circle on the shifted cylinder is mapped to a circle in the $Y$-direction at 
constant $\overline{Y}$, and the equivalence relation \eqref{eq:ec} is mapped to
\be \label{eq:gcond}
    \bigl( Y\,, \overline{Y} \bigr) : \,\, \bigl( 0\,, \overline{Y} \bigr) \sim \bigl( 2\pi r\,, \overline{Y} \bigr)\,,
    	\qquad
    r = 2 E R\,,
\ee
where $Y \in [0\,,2\pi r]$ and $\overline{Y} \in \mathbb{R}$\,. We then rewrite \eqref{eq:NROSfree2} as
\be \label{eq:snrosY}
    S_\text{NROS} = \frac{1}{4\pi\alpha'} \int_\Sigma d^2 \sigma \! \ls \p_\alpha X^{A'} \p^\alpha X^{A'} \! + \lambda' \, \overline{\p} \bigl( Y + E \, \overline{Y} \bigr) + \overline{\lambda} \, \p \overline{Y} \rs,
\ee
with the boundary conditions
\begin{subequations} \label{eq:Ybdryconds}
\begin{align}
    \p_\sigma X^{A'} \big|_{\sigma = 0} & = 0\,,
        &%
    E \, \lambda' + \overline{\lambda} \, \big|_{\sigma = 0} & = 0\,, \\[2pt]
    \overline{\p} (Y + E \, \overline{Y}) \big|_{\sigma = 0} = \p \overline{Y} \big|_{\sigma = 0} & = 0\,, 
        & %
    \p Y + E \, \overline{\p Y} \, \big|_{\sigma = 0} & = 0\,.
\end{align}
\end{subequations}
Here, the parameter $E$ controls the size of the circle on which $Y$ is compactified; the radius of this circle is $r = 2 E R$. The dispersion relation for the open strings is
\be
    \varepsilon = \frac{E}{w r} \bigl( k^{A'} k^{A'} + N_\text{open} \bigr)\,,
\ee
where $\varepsilon$ is the effective energy conjugate to $\overline{Y}$ and $k^{A'}$ denotes the transverse momentum.

Now, we perform a T-duality transformation along the $Y$-direction on the shifted cylinder. First, we introduce the parent action for \eqref{eq:snrosY},
\begin{align} \label{eq:pa}
    S_\text{parent} & = \frac{1}{4\pi\alpha'} \int_\Sigma d^2 \sigma \ls \p_\alpha X^{A'} \p^\alpha X^{A'} \! + \lambda' \bigl( \overline{v} + E \, \overline{\p Y} \bigr) + \overline{\lambda} \, \p \overline{Y} \! - \bigl( \p \tilde{Y} \, \overline{v} - \overline{\p} \tilde{Y} v \bigr) \rs, 
\end{align}
with the boundary conditions
\begin{subequations}
\begin{align}
    \p_\sigma X^{A'} \big|_{\sigma = 0} & = 0\,,
        &%
    E \, \lambda' + \overline{\lambda} \, \big|_{\sigma = 0} & = 0\,, \\[2pt]
    \overline{v} + E \, \overline{\p Y} \big|_{\sigma = 0} = \p \overline{Y} \big|_{\sigma = 0} & = 0\,, 
        & %
    v + E \, \overline{\p Y} \, \big|_{\sigma = 0} & = 0\,.
\end{align}
\end{subequations}
Integrating out $\tilde{Y}$ recovers the original action \eqref{eq:snrosY} together with the boundary conditions in \eqref{eq:Ybdryconds}. 
To perform the T-duality transformation, we instead integrate out $\overline{v}$ by plugging in the equation of motion, $\lambda' = \p \tilde{Y}$, from varying $\overline{v}$ in \eqref{eq:pa}.
We also define $\tilde{\lambda} \equiv v$\,. The resulting dual action is \cite{Gomis:2000bd} 
\be \label{eq:dualNCOS}
    \tilde{S}_\text{NCOS} = \frac{1}{4\pi\alpha'} \int_\Sigma d^2 \sigma \lr \p_\alpha X^{A'} \p^\alpha \! X^{A'} \! + E \, \p \tilde{Y} \, \overline{\p Y} + \tilde{\lambda} \, \overline{\p} \tilde{Y} + \overline{\lambda} \, \p \overline{Y} \rr,
\ee
and the dual boundary conditions are
\begin{subequations} \label{eq:dualNCOSbdry}
\begin{align}
    \p_\sigma X^{A'} \big|_{\sigma = 0} & = 0\,,
        &%
    \overline{\lambda} + E \, \p \tilde{Y} \big|_{\sigma = 0} & = 0\,, \\[2pt]
    \overline{\p} \tilde{Y} \big|_{\sigma = 0} = \p \overline{Y} \big|_{\sigma = 0} & = 0\,, 
        & %
    \tilde{\lambda} + E \, \overline{\p Y} \big|_{\sigma = 0} & = 0\,.
\end{align}
\end{subequations}
In the dual frame, the parameter $E$ is essentially a constant $B$-field and the dual coordinate $\tilde{Y}$ is compactified over a lightlike circle of radius $\tilde{r} = \alpha' / r$\,.
The dual D$(d-1)$-brane is filling in the entire spacetime.
This dual theory is the  DLCQ description of NCOS \cite{Gomis:2000bd, Danielsson:2000mu}, where $\tilde{Y}$ and $\overline{Y}$ do not commute,
\be
    [\tilde Y(\tau)\,, \overline{Y}(\tau')] = 4 \pi \alpha' \, E^{-1} \epsilon (\tau - \tau')\,.
\ee
Here, $\epsilon (\tau - \tau') = 1$ when $\tau > \tau'$ and $\epsilon (\tau - \tau') = - 1$ when $\tau < \tau'$. 

The nonrelativistic open string vertex operator \eqref{eq:osvo} in the original theory can be written in terms of $\lambda'$\,, $Y$ and $\overline{Y}$ as
\be
    \CV_\text{NROS} = \int_{\p\Sigma} d\tau \ls E N \lambda' + i \bigl( A_0 \, \p_\tau \overline{Y} + A_{A'} \, \p_\tau X^{A'} \bigr) \rs. 
\ee
The T-dual of this vertex operator gives
\begin{align} 
    \tilde{\CV}_\text{NCOS} & = i  \int_{\p\Sigma} d\tau \, \tilde{A}_\mu \, \p_\tau \tilde{Y}^\mu,
\end{align}
where $\tilde{Y}^{A'} = X^{A'}$, $\tilde{A}_{A'} = A_{A'}$ and
\begin{subequations}
\begin{align}
    \tilde{Y}^0 & = \tfrac{1}{2} \bigl( \tilde{Y} + \overline{Y} \bigr)\,, 
        \qquad
    \tilde{A}_0 = A_0 - {2} E N\,, 
    \\[2pt]
    \tilde{Y}^1 & = \tfrac{1}{2} \bigl( \tilde{Y} - \overline{Y} \bigr)\,,
        \qquad
    \tilde{A}_1 = - A_0 - {2} E N\,.
\end{align}
\end{subequations}
Here, $\tilde{A}_\mu$ defines the dual $U(1)$ gauge field.

By performing the T-duality transformation along the lightlike $\tilde{Y}$ direction in the dual NCOS theory, we will be led back to the original nonrelativistic open string theory in \eqref{eq:snrosY}. To see this, we start with the parent action that is equivalent to \eqref{eq:dualNCOS},
\be
    \tilde{S}_\text{parent} = \frac{1}{4\pi\alpha'} \int_\Sigma d^2 \sigma \ls \p_\alpha X^{A'} \p^\alpha \! X^{A'} \! + E \, u \, \overline{\p Y} + \tilde{\lambda} \, \overline{u} + \overline{\lambda} \, \p \overline{Y} - \left(\p Y \overline{u} - \overline{\p} Y u \right)\rs,
\ee
with the boundary conditions 
\begin{subequations}
\begin{align}
    \p_\sigma X^{A'} \big|_{\sigma = 0} & = 0\,,
        &%
    \overline{\lambda} + E \, u \big|_{\sigma = 0} & = 0\,, \\[2pt]
    \overline{u} \big|_{\sigma = 0} = \p \overline{Y} \big|_{\sigma = 0} & = 0\,, 
        & %
    \tilde{\lambda} + E \, \overline{\p Y} \big|_{\sigma = 0} & = 0\,.
\end{align}
\end{subequations}
By integrating out $Y$ in $\tilde{S}_\text{parent}$, we return to \eqref{eq:dualNCOS} with the boundary condition \eqref{eq:dualNCOSbdry}.  
Instead, we perform the inverse T-duality transformation by integrating out $\overline u$\,, which gives the constraint $\tilde \lambda = \p  Y$\,.
Upon plugging this constraint  back into $\tilde S_\text{parent}$ along with the identification
$\lambda' = u$ leads to the dual action that takes the same form as the action \eqref{eq:snrosY}. The dual boundary conditions also coincide with the ones in \eqref{eq:Ybdryconds}.
  
Note that, for the dual theory to be NCOS, it is essential to take $L = 2 \pi R$ (or $L = - 2 \pi R$) for the shifted cylinder defined by the gluing condition \eqref{eq:ec}. For any other value of $L \neq \pm 2 \pi R$, the dual theory is instead the DLCQ of relativistic open strings coupled to a constant $B$-field. This latter case can be shown straightforwardly by following the same procedure in \S\ref{sec:relnonrel}, which we brief as follows. Define $s = L / (2 \pi R) \neq \pm 1$\,, then the change of variable in \eqref{eq:cvlp} becomes
\be \label{eq:cvlp2}
    \lambda' = E^{-1} \lambda\,,
        \qquad
    Y = 2 E X^1,
        \qquad
    \overline{Y} = X^0 - s \, X^1,
\ee
such that the gluing condition \eqref{eq:gcond} remains the same for $(Y, \overline{Y})$. In terms of $Y$ and $\overline{Y}$, the action \eqref{eq:NROSfree2} becomes
\begin{align}
    S_\text{NROS} 
    & = \frac{1}{4\pi\alpha'} \int_\Sigma d^2 \sigma \! \lc \p_\alpha X^{A'} \p^\alpha X^{A'} \! + \lambda' \, \overline{\p} \lr \tfrac{s+1}{2} \, Y + E \, \overline{Y} \rr + \overline{\lambda} \, \p \! \lr \tfrac{s-1}{2 E} \, Y + \overline{Y} \rr \rc.
\end{align}
Performing the T-duality transformation as in \S\ref{sec:relnonrel} along the circular direction $Y$, we find that the dual action takes the form
\begin{align}
    S_\text{ROS} 
     & = \frac{1}{4\pi\alpha'} \int_\Sigma d^2 \sigma \ls \p_\alpha X^{A'} \p^\alpha \! X^{A'} \! + \tfrac{4E}{s^2 - 1} \, \bigl( \p_\alpha \tilde{Y} \, \p^\alpha \overline{Y} + i \, s \, \epsilon^{\alpha\beta} \p_\alpha \tilde{Y} \, \p_\beta \overline{Y} \bigr) \rs,
\end{align}
with the boundary conditions 
\be
	\p_\sigma X^{A'} \big|_{\sigma=0} = 0\,, 
		\qquad
	\p_\sigma \tilde{Y} \big|_{\sigma=0} = - i \, s \, \p_\tau \tilde{Y}\,,
		\qquad
	\p_\sigma \overline{Y} \big|_{\sigma = 0} = i \, s \, \p_\tau \overline{Y}\,.
\ee 
This is relativistic open string theory with a lightlike circle along the $\tilde{Y}$ direction and in a constant $B$-field proportional to $4 E s / (s^2 - 1)$\,.

\subsection{From Noncommutative to Relativistic Open Strings} \label{sec:noncommtorel}

We have so far studied the T-duality transformation that maps from nonrelativistic open string theory to the DLCQ of relativistic and noncommutative open string theory, respectively. For completeness   we perform in this last subsection a longitudinal spatial T-duality transformation in NCOS, whose dual gives rise to relativistic open string theory with a D($d-2$)-brane in the DLCQ description. In the dual relativistic open string theory, the D($d-2$)-brane moves in the dual circle with a velocity   approaching the speed of light, and thus with an infinite momentum. The following discussion provides a first principles derivation of the T-duality transformation between NCOS and the DLCQ of relativistic string theory, which was previously studied, for example, in \cite{Seiberg:2000ms} as a subtle limit.

We start with NCOS as defined in \cite{Gomis:2000bd}, or in \eqref{eq:dualNCOS}, 
\be \label{eq:oriNCOS}
    S_\text{NCOS} = \frac{1}{4\pi\alpha'} \int_\Sigma d^2 \sigma \bigl( \p_\alpha X^{A'} \p^\alpha X^{A'} + E \, \p X \, \overline{\p X} + \lambda \, \overline{\p} X + \overline{\lambda} \, \p \overline X \bigr)\,,
\ee
with $E$ a constant $B$-field and the boundary conditions in \eqref{eq:dualNCOSbdry},
\begin{subequations} \label{eq:bdryNCOStoROS}
\begin{align}
    \p_\sigma X^{A'} \big|_{\sigma = 0} & = 0\,,
        &%
    \overline{\lambda} + E \, \p X \big|_{\sigma = 0} = 0\,, \\[2pt]
    \overline{\p} X \big|_{\sigma = 0} = \p \overline{X} \big|_{\sigma = 0} & = 0\,, 
        & %
    {\lambda} + E \, \overline{\p X} \big|_{\sigma = 0} = 0\,.
\end{align}
\end{subequations}
Recall that this set of boundary conditions imply that the theory lives on a spacetime-filling D$(d-1)$-brane.

To perform a longitudinal spatial T-duality transformation along the $X^1$ direction compactified over a circle of radius $R$\,, we start with the parent action 
\begin{align} \label{eq:NCOSparent}
	S_\text{parent} & = \frac{1}{4\pi\alpha'} \int_\Sigma d^2 \sigma \, \Bigl[ \p_\alpha X^{A'} \p^\alpha \! X^{A'} + E \bigl( \p X^0 + v \bigr) \bigl( \overline{\p} X^0 - \overline{v} \bigr) \Bigr] \notag \\[2pt]
	& \quad + \frac{1}{4\pi\alpha'} \int_\Sigma d^2 \sigma \, \Bigl[ \lambda \, \bigl( \overline{\p} X^0 + \overline{v} \bigr) + \overline{\lambda} \, \bigl( \p X^0 - v  \bigr) \Bigr] \notag \\[2pt]
	& \quad + \frac{i}{2\pi\alpha'} \int_{\Sigma} d^2 \sigma \, \epsilon^{\alpha\beta} \, \tilde{X} \, \p_\alpha v_\beta + \frac{i}{2\pi\alpha'} \int_{\p\Sigma} d\tau \, \tilde{X} \, v_\tau\,,
\end{align}
with $v = v_\sigma - i v_\tau$\,, $\overline{v} = v_\sigma + i v_\tau$\,, and the boundary conditions
\begin{subequations} \label{eq:bdryNCOStoROSp}
\begin{align}
    \p_\sigma X^{A'} \big|_{\sigma = 0} & = 0\,,
        &%
    \overline{\lambda} + E \, \bigl( \p X^0 + v \bigr) \big|_{\sigma = 0} & = 0\, \\[2pt]
    \overline{\p} X^0 + \overline{v} \big|_{\sigma = 0} = \p X^0 - v \big|_{\sigma = 0} & = 0\,, 
        & %
    {\lambda} + E \, \bigl( \overline{\p} X^0 - \overline{v} \bigr) \big|_{\sigma = 0} & = 0\,.
\end{align}
\end{subequations}
The global translational symmetry in $\tilde{X}$ requires introducing the boundary term in \eqref{eq:NCOSparent}.
We also require the Dirichlet boundary condition $\delta \tilde{X} \big|_{\sigma = 0} = 0$\,. Then, integrating out $\tilde{X}$ gives back the original NCOS theory in \eqref{eq:oriNCOS}. 

To proceed with the longitudinal T-duality transformation in NCOS, we integrate out $v_\alpha$ in \eqref{eq:NCOSparent} by imposing the equations of motion from varying with respect to $v_\alpha$\,,
\begin{subequations} \label{eq:vrules}
\begin{align}
     v_\tau & = \tfrac{1}{2} i E^{-1} \bigl( \lambda + \overline{\lambda} \bigr) -i \, \p_\sigma \bigl( X^0 - E^{-1} \tilde{X} \bigr)\,, \\[2pt]
    v_\sigma & = \tfrac{1}{2} E^{-1} \bigl( \lambda - \overline{\lambda} \bigr) + i \, \p_\tau \bigl( X^0 - E^{-1} \tilde{X} \bigr)\,.
\end{align}
\end{subequations}
Subsequently, we integrate out $\lambda$ and $\overline{\lambda}$ in the resulting action, by plugging in their equations of motion,
\begin{align} \label{eq:lambdarules}
    \lambda = \p \bigl( 2 E X^0 - \tilde{X} \bigr)\,,
        \qquad
    \overline{\lambda} = \overline{\p} \bigl( 2 E X^0 - \tilde{X} \bigr)\,.
\end{align}
Note that \eqref{eq:vrules} and \eqref{eq:lambdarules} are consistent with the equations of motion
\be \label{eq:vtausigmaNCOS}
    v_\tau = i \, \p_\sigma X^0,
        \qquad
    v_\sigma = - i \, \p_\tau X^0,
\ee
imposed by the Lagrange multipliers $\lambda$ and $\overline{\lambda}$ in \eqref{eq:NCOSparent}.
We find the dual action
\be \label{eq:dualROSfNCOS}
    \tilde{S}_\text{ROS} = \frac{1}{4\pi\alpha'} \int d^2 \sigma \lr \p_\alpha X^{A'} \, \p^\alpha \! X^{A'} + 4 \, E \, \p_\alpha X^0 \, \p^\alpha X^0 - 2 \, \p_\alpha X^0 \, \p^\alpha \tilde{X} \rr,
\ee
together with the dual boundary conditions
\be \label{eq:bdryROSfNC}
    \p_\tau \tilde{X} \big|_{\sigma = 0} = 0\,, 
        \qquad
    \p_\sigma X^{A'} \big|_{\sigma = 0} = 0\,,
        \qquad
    \p_\sigma \bigl( \tilde{X} - 4 \, E \, X^0 \bigr) \big|_{\sigma = 0} = 0\,.
\ee
In this dual theory, $\tilde{X}$ is compatified on a lightlike circle of radius $\tilde{R} = \alpha' / R$\,. This is relativistic string theory in the DLCQ description. 
The Dirichlet boundary condition in the lightlike $\tilde{X}$ direction implies that there is a D($d-2$)-brane boosted to large momentum along the $\tilde{X}$ circle \cite{Seiberg:2000ms}.     

The most general open string vertex operator in NCOS is
\be
    \CV_\text{NCOS} = i \int_{\p\Sigma} d\tau \, A_\mu \, \p_\tau X^\mu,
\ee
which in the parent theory becomes
\be
    \CV_\text{parent} = i \int_{\p\Sigma} d\tau \, \bigl( A_0 \, \p_\tau X^0 + A_1 \, v_\tau + A_{A'} \p_\tau X^{A'} \bigr)\,.
\ee
Applying the duality map \eqref{eq:vtausigmaNCOS} and the last boundary condition in \eqref{eq:bdryROSfNC}, we find the dual vertex operator,
\be
    \tilde{\CV}_\text{ROS} = \int_{\p\Sigma} d\tau \, \bigl[ \tilde{A} \, \p_\sigma \tilde{X} + i \bigl( A_0 \, \p_\tau X^0 + A_{A'} \p_\tau X^{A'} \bigr) \bigr]\,,
\ee
where $\tilde{A} = - A_1 / (4 E)$\,.

To perform the inverse T-duality transformation, we consider the following parent action that is equivalent to \eqref{eq:dualROSfNCOS}:
\begin{align}
    \tilde{S}_\text{parent} & = \frac{1}{4\pi\alpha'} \int d^2 \sigma \lr \p_\alpha X^{A'} \, \p^\alpha \! X^{A'} + 4 E \, \p_\alpha X^0 \, \p^\alpha X^0 - 2 \, \p^\alpha X^0 \, u_\alpha \rr \notag \\[2pt]
    & \quad + \frac{i}{2\pi\alpha'} \int_\Sigma d^2 \sigma \, \epsilon^{\alpha\beta} \, X^1 \p_\alpha u_\beta\,,
\end{align}
with the boundary conditions
\be 
    u_\tau \big|_{\sigma = 0} = 0\,, 
        \qquad
    \p_\sigma X^{A'} \big|_{\sigma = 0} = 0\,,
        \qquad
    u_\sigma - 4 E \, \p_\sigma X^0 \big|_{\sigma = 0} = 0\,.
\ee
Upon integrating out $X^1$ we return to the original action \eqref{eq:dualROSfNCOS}. Applying the change of variables $\lambda' = - u_\sigma + i u_\tau$ and $\overline{\lambda}{}' = - u_\sigma - i u_\tau$, we find the dual action 
\be \label{eq:Slambdap}
    S_\text{NCOS} = \frac{1}{4\pi\alpha'} \int d^2 \sigma \lr \p_\alpha X^{A'} \, \p^\alpha \! X^{A'} + 4 E \, \p X^0 \, \overline{\p} X^0 + \lambda' \, \overline{\p} X + \overline{\lambda}{}' \, \p \overline{X} \rr,  
\ee
together with the boundary conditions
\begin{align}
    \p_\sigma X^{A'} \big|_{\sigma = 0} & = 0\,,
        \qquad%
    \lambda' - \overline{\lambda}{}' \, \big|_{\sigma = 0} = 0\,, 
        \qquad%
    \lambda' + \overline{\lambda}{}' \, \big|_{\sigma = 0} = - 8 E \, \p_\sigma X^0 \big|_{\sigma = 0}\,.
\end{align}
Identifying,
\be
    \lambda = \lambda' + E \, \p \bigl( X + \tfrac{1}{2}  \overline{X} \bigr)\,,
        \qquad
    \overline{\lambda} = \overline{\lambda}{}' + E \, \overline{\p} \bigl( \tfrac{1}{2} X + \overline{X} \bigr)\,,
\ee
we find that the dual action \eqref{eq:Slambdap} and the associated boundary conditions are indeed identical to \eqref{eq:oriNCOS} and \eqref{eq:bdryNCOStoROS}, respectively.

\section{Curved Spacetime} \label{sec:cs}

In this section, we generalize the results in \S\ref{sec:flat} to sigma models in arbitrary closed and open string background fields. The calculation is most straightforward if we perform the T-duality transformations on the side of the theory with a spacetime-filling brane.  In this case we can absorb the open string $U(1)$ gauge field on the brane into the definition of the $B$-field and then directly apply the closed string Buscher rules that are already derived in \cite{Bergshoeff:2018yvt, Bergshoeff:2019pij}, following the procedure introduced in \cite{Alvarez:1996up}. In each of the following subsections, we will first give a brief derivation for each T-duality transformation that leads to the same set of Buscher rules in \cite{Bergshoeff:2018yvt, Bergshoeff:2019pij}. The derivations here will be done in conformal gauge, without introducing the dilaton field. This makes the T-duality transformations simpler to handle compared to the complete derivation with the dilaton field in \cite{Bergshoeff:2018yvt}. Revisiting how the T-duality transformation is performed allows us to extract the duality relations that are needed to take the T-dual of the boundary conditions. For the T-duality transformation of the dilaton field, we refer to the result in \cite{Bergshoeff:2018yvt}. We will discuss the T-duality transformations on the spacetime Dirac-Born-Infeld type actions at the end of each subsection.

\subsection{From Relativistic to Nonrelativistic Open Strings} \label{sec:RtoNR}

We start with generalizing the results in \S\ref{sec:relnonrel} that maps between relativistic and nonrelativistic open strings to arbitrary background fields.
Consider the sigma model of (relativistic) string theory on a $d$-dimensional arbitrary curved Riemannian geometry and $B$-field background, in presence of a D$(d-1)$-brane that is spacetime filling,
\begin{align} \label{eq:relaction}
    S_\text{ROS} & = \frac{1}{4\pi\alpha'} \int_\Sigma d^2 \sigma \Bigl( \p_\alpha X^\mu \, \p^\alpha X^\nu \, g_{\mu\nu} - i \, \epsilon^{\alpha\beta} \, \p_\alpha X^\mu \, \p_\beta X^\nu \, b_{\mu\nu} \Bigr) \notag \\[2pt]
    & \quad + \frac{i}{2\pi\alpha'} \int_{\p\Sigma} d\tau \, A_\mu \, \p_\tau X^\mu \,.
\end{align}
Here, we denote the metric field by $g_{\mu\nu}$ and the Kalb-Ramond field by $b_{\mu\nu}$\,. The $U(1)$ gauge field $A_\mu$ denotes the massless degrees of freedom living on the D-brane. 
We disregarded the dilaton field, which plays the same role as in closed string theory and can be reintroduced easily at the end of the calculation. The boundary of the worldsheet is at $\sigma = 0$\,. The Neumann boundary condition at $\sigma = 0$ is
\be \label{eq:firstbdrycond}
    g_{\mu\nu} \p_\sigma X^\nu + i \bigl( b_{\mu\nu} + F_{\mu\nu} \bigr) \p_\tau X^\nu = 0\,,
\ee
where $F_{\mu\nu} = \p_\mu A_\nu - \p_\nu A_\mu$\,.  If the open string is neutral under the background $U(1)$ gauge field on the D-brane, we can lift the boundary field $A_\mu$ into the bulk.  This allows us to read off the T-duality transformation of the boundary fields directly by using the closed string Buscher rules, followed by restricting $A_\mu$ to be a boundary field at the end of the calculation.  We thus rewrite the action \eqref{eq:relaction} as
\begin{align} \label{eq:relaction1}
    S_\text{ROS} & = \frac{1}{4\pi\alpha'} \int_\Sigma d^2 \sigma \Bigl( \p_\alpha X^\mu \, \p^\alpha X^\nu \, g_{\mu\nu} - i \, \epsilon^{\alpha\beta} \, \p_\alpha X^\mu \, \p_\beta X^\nu \, \CB_{\mu\nu} \Bigr)\,,
\end{align}
with $\CB_{\mu\nu} = b_{\mu\nu} + F_{\mu\nu}$\,.  We assume that there exists a lightlike Killing vector $k^\mu$\,, with
\be \label{eq:lightlikecond}
    g_{\mu\nu} \, k^\mu \, k^\nu = 0\,.
\ee 
We introduce a coordinate system $X^\mu = (y\,, X^i)$ adapted to $k^\mu$\,, such that $k^\mu \p_\mu = \p_y$\,. The action is then invariant under this target space isometry provided that
\be
	\CL_k \, g_{\mu\nu} = \CL_k \, \CB_{\mu\nu} = 0\,,
\ee
where $\CL_k$ is the Lie derivative in the direction $k^\mu$.
To perform a T-duality transformation along the isometry ${y}$-direction, we now use the formulation developed by Ro\v{c}ek and Verlinde in \cite{Rocek:1991ps}, which has the benefit of preserving the general covariance and manifesting effects of the global topology of the dual manifold; it is also easier to generalize this method to more complicated cases.

The abelian isometry is given by $\delta_\epsilon y = \epsilon$\,.  We will gauge this isometry by introducing an auxiliary gauge field $v_\alpha$ that transforms as $\delta_\epsilon v_\alpha = - \p_\alpha \epsilon$\,.  The gauged version of the string action \eqref{eq:relaction1} is 
\begin{align} \label{eq:parentaction}
    S_\text{gauged} & = \frac{1}{4\pi\alpha'} \int_\Sigma d^2 \sigma \Bigl( D_\alpha X^\mu \, D^\alpha X^\nu \, g_{\mu\nu} - i \, \epsilon^{\alpha\beta} D_\alpha X^\mu \, D_\beta X^\nu \, \CB_{\mu\nu} \Bigr) \notag \\[2pt]
    & \quad - \frac{i}{2\pi\alpha'} \int_\Sigma d^2 \sigma \, \epsilon^{\alpha\beta} \, \tilde{y} \, \p_\alpha v_\beta - \frac{i}{2\pi\alpha'} \int_{\p\Sigma} d\tau \, \tilde{y} \, v_\tau \,,
\end{align}
where we introduced a covariant derivative $D_\alpha$ with $D_\alpha X^\mu = \p_\alpha X^\mu + k^\mu \, v_\alpha$\,. Furthermore, we require the Dirichlet boundary condition 
\begin{equation} \label{eq:Dty}
    \tilde y \, \big|_{\sigma=0} = \tilde y_0
\end{equation}
for constant $\tilde{y}_0$\,, which is necessary for maintaining the gauge invariance. 
Note that the boundary term in \eqref{eq:parentaction} is required by the global translational symmetry $\delta_\xi \, \tilde{y} = \xi$\,, maintaining the physical equivalence for wherever the dual brane sits along $\tilde{y}$\,. Then, the Lagrange multiplier $\tilde y$ imposes that $v_\alpha$ has zero field strength, $\epsilon^{\alpha\beta} \p_\alpha v_\beta = 0$\,. Note that $\tilde y$ will be interpreted as the isometry direction dual to $y$. We also have the analogue of the Neumann boundary condition \eqref{eq:firstbdrycond} for the gauged action, 
\be \label{eq:Nbdrycond}
	g_{\mu\nu} D_\sigma X^\nu + i \, \CB_{\mu\nu} \, D_\tau X^\nu = 0\,.
\ee 
Fixing the gauge $v_\alpha=0$\,, the original theory in \eqref{eq:relaction1} with the boundary condition \eqref{eq:firstbdrycond} is recovered.

To pass on to the dual action, one may proceed with the same derivation as in \cite{Bergshoeff:2018yvt, Bergshoeff:2019pij} to find the dual action. However, in the absence of the dilaton field, there is a simpler method one can use to derive the same result, which we present now.

We first note that the gauged action \eqref{eq:parentaction} is linear in $v_\alpha$\,, which is a Lagrange multiplier that imposes the condition
\begin{align} \label{eq:lambdaconstraints}
	g_{yi} \, \p_\alpha X^i - i \, \epsilon_\alpha{}^\beta \bigl[ \bigl( b_{yi} - \p_i A_y \bigr) \, \p_\beta X^i + \p_\beta \tilde{y} \, \bigr] = 0\,.
\end{align}
Moreover, we define the dual coordinates $\tilde X^\mu = (\tilde y , X^i)$ adapted to the dual isometry direction, and also the covariant derivative
\be \label{dualCovDerivative}
	D_\alpha \tilde{X}^\mu = \p_\alpha \tilde{X}^\mu - k^\mu \, \p_\alpha X^i \, \p_i A_y\,,
		\qquad
	k^\mu \p_\mu = \p_{\tilde{y}}\,,
\ee
as required by the Stueckelberg symmetry
\be \label{eq:deltaeta}
	\delta_\eta \tilde{y} = \eta (X^i)\,,
		\qquad
	\delta_\eta A_y = \eta (X^i)\,,
\ee
which gauges the global translational symmetry in $\tilde{y}$\,. 

In terms of $\tilde{X}^\mu$, the parent action \eqref{eq:parentaction} becomes
\begin{align} \label{eq:Sguaged2}
    S_\text{gauged} & = \frac{1}{4\pi\alpha'} \int_\Sigma d^2 \sigma \Bigl( \p_\alpha X^i \, \p^\alpha X^j g_{ij} - i \, \epsilon^{\alpha\beta} \p_\alpha X^i \, \p_\beta X^j \, b_{ij} \Bigr) \notag \\[2pt]
    & \quad + \frac{1}{4\pi\alpha'} \int_\Sigma d^2 \sigma \Bigl\{ \bigl( v + \p \, \tilde{y} \, \bigr) \, \Bigl[  \overline{\p} X^i \, \bigl( g_{yi} + b_{yi} \bigr) + \bigl( \overline{\p} \, \tilde{y} - \overline{\p} X^i \, \p_i A_y \bigr) \Bigr] \notag \\[2pt]
	& \hspace{2.85cm} + \bigl( \overline{v} + \overline{\p} \, \tilde{y} \, \bigr) \, \Bigl[ \p X^i \, \bigl( g_{yi} - b_{yi} \bigr) - \bigl( \p \, \tilde{y} - \p X^i \, \p_i A_y \bigr) \Bigr] \Bigr\} \notag \\[2pt]
	& \quad + \frac{i}{2\pi\alpha'} \int_{\p\Sigma} d\tau \, \bigl( A_i \, \p_\tau X^i + \tilde{y}_0 \, \p_\tau y \bigr)\,,
\end{align}
where
\begin{subequations} \label{eq:lambdarel0}
\begin{align}
	\p & = \p_\sigma - i \, \p_\tau\,,
		\qquad 
	v = v_\sigma - i v_\tau\,, \\[2pt]
	\overline{\p} & = \p_\sigma + i \, \p_\tau\,,
		\qquad 
	\overline{v} = v_\sigma + i v_\tau\,.
\end{align}
\end{subequations}
Upon the identification 
\be \label{eq:lambda'v}
	\lambda' = v + \p y\,,
		\qquad
	\overline{\lambda}{}' = \overline{v} + \overline{\p} y\,,
\ee
the bulk action in \eqref{eq:lambdarel0} already takes the form of nonrelativistic string theory in \cite{Bergshoeff:2018yvt}. Note that \eqref{eq:lambda'v} is essentially the gauge invariant version of \eqref{eq:lambdau} that we already encountered in the discussion for flat spacetime.

In nonrelativistic string theory, there is a Stueckelberg symmetry\,\footnote{This should not be confused with the Stueckelberg transformation in \eqref{eq:deltaeta} from gauging the isometry $\tilde{y}$\,.} that mixes the target space geometry and the $B$-field \cite{Bergshoeff:2018yvt}, which can be introduced in \eqref{eq:Sguaged2} by taking the change of variables
\be \label{eq:deflambda}
	\lambda = C^{-1} \ls \, \lambda' - D \tilde{X}^\mu \bigl( C_\mu{}^0 - C_\mu{}^1 \bigr) \rs,
		\qquad
	\overline{\lambda} = C^{-1} \ls \, \overline{\lambda}{}' - \overline{D} \tilde{X}^\mu \bigl( C_\mu{}^0 + C_\mu{}^1 \bigr) \rs,
\ee
where we defined
\be
	D = D_\sigma - i D_\tau\,, 
		\qquad 
	\overline{D} = D_\sigma + i D_\tau\,.
\ee
Moreover, $C = C(\tilde{X}) \neq 0$ and $C_\mu{}^A = C_\mu{}^A (\tilde{X})$ are arbitrary functions that parametrize the Stueckelberg transformations that mix the dual target space geometry and $B$-field in the dual nonrelativistic string theory \cite{Bergshoeff:2018yvt}. The dependence on $C$ and $C_\mu{}^A$ in \eqref{eq:deflambda} is introduced such that it matches the Stueckelberg transformations of $\lambda$ and $\overline{\lambda}$ in \cite{Bergshoeff:2018yvt} (also see \eqref{eq:Slambda} and \eqref{eq:Slambdabar}), but with the derivatives on ${\tilde{X}}^\mu$ gauged as in \eqref{dualCovDerivative}. We will come back to this Stueckelberg symmetry later around \eqref{eq:unbrokensymmetries4}.

The dual action can then be obtained by rewriting \eqref{eq:Sguaged2} (together with the boundary condition \eqref{eq:Nbdrycond}) by substituting $\lambda', \overline{\lambda}{}'$ with $\lambda\,, \overline{\lambda}$ using \eqref{eq:deflambda}. The dual action is
\begin{align} \label{eq:dualactionstandartform0}
    \tilde{S}_\text{NROS} & = \frac{1}{4\pi\alpha'} \int_\Sigma d^2 \sigma \Bigl( D_\alpha \tilde{X}^\mu D^\alpha \tilde{X}^\nu H_{\mu\nu} - i \, \epsilon^{\alpha\beta} D_\alpha \tilde{X}^\mu D_\beta \tilde{X}^\nu B_{\mu\nu} \Bigr) \notag \\[2pt]
    & \quad + \frac{1}{4\pi\alpha'} \int_\Sigma d^2 \sigma \Bigl( \lambda \, \overline{D} \tilde{X}^\mu \, \tau_\mu + \overline{\lambda} \, D \tilde{X}^\mu \, \overline{\tau}_\mu \Bigr) \notag \\[2pt]
	& \quad + \frac{i}{2\pi\alpha'} \int_{\p\Sigma} d\tau \, \bigl( A_i \, \p_\tau X^i + \tilde{y}_0 \, \p_\tau y \bigr)\,, 
\end{align}
where
%
		%
%
%
\begin{subequations} \label{eq:HBAaction}
\begin{align}
    \tau^{}_i{}^0 & = C \, g_{yi}\,, 
        \quad\,\qquad%
    \tau_y{}^0 = 0\,, \\[2pt]
    \tau^{}_i{}^1 & = C\, b_{yi}\,, 
        \quad\,\qquad%
    \tau_y{}^1 = C\,, \\[2pt]
    H_{yy} & = - 2 \, C_y{}^1\,, 
        \qquad%
    H_{yi} = C_{y}\,^0 \, g_{yi} - C_y\,^1 \, b_{yi} - C_i{}^1\,, \\[2pt]
    B_{yi} & = C_y{}^0 \, b_{yi} - C_y{}^1 \, g_{yi} - C_i{}^0\,, \\[2pt]
    H_{ij} & = g_{ij} + \bigl( C_i{}^0 \, g_{yj} + C_j{}^0 \, g_{yi} \bigr) - \bigl( C_i{}^1 \, b_{yj} + C_j{}^1 \, b_{yi} \bigr)\,, \\[2pt]
    B_{ij} & = b_{ij} + \bigl( C_i{}^0 \, b_{yj} - C_j{}^0 \, b_{yi} \bigr) - \bigl( C_i{}^1 \, g_{yj} - C_j{}^1 \, g_{yi} \bigr)\,.
\end{align}
\end{subequations}
This bulk action is precisely nonrelativistic string theory on a string Newton-Cartan geometry and $B$-field background that generalizes the free action \eqref{eq:NROSfree} in flat spacetime \cite{Bergshoeff:2018yvt}. Note that $\tau_\mu{}^A$ plays the role of the longitudinal Vielbein field, with $\tau_\mu = \tau_\mu{}^0 + \tau_\mu{}^1$ and $\overline{\tau}_\mu = \tau_\mu{}^0 - \tau_\mu{}^1$. Define $E_\mu{}^{A'}$ to be the transverse Vielbein field in string Newton-Cartan geometry, we write as in \cite{Bergshoeff:2018yvt}
\be
    H_{\mu\nu} = E_\mu{}^{A'} E_\nu{}^{A'} + \bigl( \tau_\mu{}^A m_\nu{}^B + \tau_\nu{}^A m_\mu{}^B \bigr) \, \eta_{AB}\,,
\ee
where $m_\mu{}^A$ is the gauge field associated with a noncentral extension in the string Newton-Cartan algebra.
The Buscher rules in \eqref{eq:HBAaction} are the same as in \cite{Bergshoeff:2019pij}. The dilaton transformation does not change, which can be read off from \cite{Bergshoeff:2019pij}, with
\be \label{eq:dilaton}
	\Phi = \phi + \log |C|\,.
\ee
Here, $\phi$ is the dilaton in the original theory and $\Phi$ is the dual dilaton. We thus recover the result that the dual target space is described by string Newton-Cartan geometry, coupled to the dual $B$-field (and dilaton). 
Moreover, we see that the constraint \eqref{eq:lambdaconstraints} imposed by integrating out $v_\alpha$ in \eqref{eq:parentaction} can be equivalently written as
\be \label{eq:constraintsnew}
    \overline{D} \tilde{X}^\mu \, \tau_\mu = D \tilde{X}^\mu \, \overline{\tau}_\mu = 0\,,
\ee
which are imposed by the Lagrange multipliers $\lambda$ and $\overline{\lambda}$ in the dual theory \eqref{eq:dualactionstandartform0}.

When $y$ is compactified on a circle of radius $R$\,, the boundary term in \eqref{eq:dualactionstandartform0} gives 
\be
	\oint_{\p\Sigma} d\tau \, \tilde{y}_0 \, \p_\tau y = 2 \pi n R \, \tilde{y}_0\,, \quad n \in \mathbb{Z}\,.
\ee
In the path integral, this gives rise to the dual of the Wilson line for the $U(1)$ gauge field, in the sector with winding number $n$. Here, the location $\tilde{y} = \tilde{y}_0$ of the D-brane is dual to the constant value that a gauge-fixed $A_y$ takes. This observation generalizes straightforwardly to the $U(1)^N$ case with multiple D-branes in the dual theory \cite{Alvarez:1996up}. 

The dual boundary conditions can be derived by using \eqref{eq:lambdaconstraints} and \eqref{eq:deflambda} to rewrite \eqref{eq:Nbdrycond} in terms of $\lambda$ and $\overline{\lambda}$\,, which gives
\begin{subequations} \label{eq:dualboundary}
\begin{align}
	\p_\tau \tilde{y} & = 0\,, \label{eq:bdrypty} \\[2pt]
	D_i \tilde{X}^\mu \ls H_{\mu\nu} \, D_{\!\sigma} \tilde{X}^\nu + i \, B_{\mu\nu} \, D_\tau \tilde{X}^\nu + \tfrac{1}{2} \bigl( \lambda \, \tau_\mu + \overline{\lambda} \, \overline{\tau}_\mu \bigr) \rs + i \, F_{ij} \, \p_\tau X^j & = 0\,,\label{eq:bdryNAy}
\end{align}
\end{subequations}
where $D_i \tilde{X}^\mu = \delta^\mu_i - k^\mu \p_i A_y$\,. Equation \eqref{eq:bdrypty} implies that $\tilde{y}$ satisfies a Dirichlet boundary condition. This is consistent with \eqref{eq:Dty} from requiring that the action \eqref{eq:parentaction} is invariant under the global translation in $\tilde{y}$\,. One also obtains the same boundary condition \eqref{eq:bdryNAy} from varying the action in \eqref{eq:dualactionstandartform0} with respect to $X^i$\,. The dual isometry $\tilde{y}\,$-direction is spacelike in the longitudinal sector, with 
$\p_{\tilde{y}} = k^\mu \p / \p {\tilde{X}}^\mu$
and 
\be
    k^\mu \, \tau_\mu{}^0 = 0\,,
        \qquad
    k^\mu \, \tau_\mu{}^1 \neq 0\,,
        \qquad
    k^\mu \, E_\mu{}^{A'} = 0\,.
\ee
In this dual theory, there is a D($d-2$)-brane localized at $\tilde{y} = \tilde{y}_0$ and transverse to the isometry $\tilde{y}\,$-direction. The global translational symmetry in $\tilde y$ is now gauged and thus spontaneously broken in the dual theory.

In the dual nonrelativistic open string theory, the field $A_y$ plays the role of a Nambu-Goldstone boson associated with the spontaneous symmetry breaking of the isometry in $\tilde{y}$\,. This can be made manifest by defining the collective coordinate $\tilde Y = \tilde{y} - A_y$\,.
Using the adapted coordinates $\tilde Y^\mu = (\tilde Y,  X^i)$\,, we rewrite \eqref{eq:dualactionstandartform0} as
\begin{align} \label{eq:unbroken}
    \tilde{S}_\text{NROS} & = \frac{1}{4\pi\alpha'} \int d^2 \sigma \Bigl( \p_\alpha \tilde Y^\mu \p^\alpha \tilde Y^\nu H_{\mu\nu} - i \, \epsilon^{\alpha\beta} \p_\alpha \tilde Y^\mu \p_\beta \tilde Y^\nu B_{\mu\nu} + \lambda \, \overline{\p} \tilde Y^\mu \, \tau_\mu + \overline{\lambda} \, \p \tilde Y^\mu \, \overline{\tau}_\mu \Bigr) \notag \\[2pt]
	& + \frac{i}{2\pi\alpha'} \int_{\p\Sigma} d\tau \, \bigl( A_i \, \p_\tau X^i + \tilde{y}_0 \, \p_\tau y \bigr)\,, 
\end{align}
and the constraints in \eqref{eq:constraintsnew} become
\be \label{eq:ahconds}
    \overline{\p} \, \tilde Y^\mu \, \tau_\mu = \p \, \tilde Y^\mu \, \overline{\tau}_\mu = 0\,.
\ee
The boundary conditions in \eqref{eq:dualboundary} become
\begin{subequations} \label{eq:unbrokenbdryconds}
\begin{align}
	\p_\tau \tilde Y + \p_\tau X^i \, \p_i A_y & = 0\,, \\[2pt]
	\p_i \tilde Y^\mu H_{\mu\nu} \, \p_{\sigma} \tilde Y^\nu + i \, \bigl( \p_i \tilde Y^\mu B_{\mu\nu} \p_j \tilde Y^\nu + F_{ij} \bigr) \, \p_\tau X^j + \tfrac{1}{2} \bigl( \lambda \, \tau_\mu + \overline{\lambda} \, \overline{\tau}_\mu \bigr) \, \p_i \tilde Y^\mu & = 0\,.
\end{align}
\end{subequations}
In the above, we are taking $A_y$ as a field living in the bulk, and we may treat it as such when performing T-duality.  Restricting $A_y$ back to be on the boundary imposes the boundary condition $\tilde Y \big|_{\sigma = 0} = \tilde y_0 - A_y$\,,
while $\tilde{Y} = \tilde y$ in the bulk.  

The Nambu-Goldstone boson $A_y$ can also be treated of as part of the embedding function $\tilde Y^\mu = f^\mu(X^i)$ that describes how the D-brane is embedded in a $d$-dimensional spacetime. We have 
\be
	\delta \tilde Y^\mu \Big|_{\sigma=0} = \delta X^i \,\p_i f^\mu(X^i).
\ee
Here, $X^i$, $i = 0,1,\ldots , d-2$\,, are coordinates of the D-brane submanifold. Varying the action in \eqref{eq:unbroken} with respect to $X^i$ gives 
\begin{align}
    & \p_i f^\mu H_{\mu\nu} \, \p_\sigma \tilde Y^\nu + i \, \CF_{ij} \, \p_\tau X^j + \tfrac{1}{2} \lr \lambda \, \tau_i +  \overline{\lambda} \, \overline{\tau}_i \rr = 0\,,
\end{align}
where
\be
    \tau_i{}^A = \p_i f^\mu \, \tau_\mu{}^A\,,
        \qquad
    \CF_{ij} = \p_i f^\mu \, B_{\mu\nu} \, \p_j f^\nu + F_{ij}\,.
\ee
The above boundary conditions (together with the equations from restricting \eqref{eq:ahconds} on the boundary) recover the ones in \cite{Gomis:2020fui} for nonrelativistic open string theory.

When $f^\mu$ is left arbitrary, the configuration of the D-brane is not fixed and we are in the unbroken phase of the spontaneous symmetry breaking of the longitudinal spatial isometry. 
The theory in \eqref{eq:unbroken} is invariant under the string Newton-Cartan gauge symmetry (except for the ones broken by the isometry direction) as well as the following Stueckelberg transformations \cite{Gomis:2020fui}:
\begin{subequations} \label{eq:unbrokensymmetries4}
\begin{align} 
    H_{\mu\nu} & \rightarrow H_{\mu\nu} - \bigl( \tau_\mu{}^A \, C_\nu{}^B + \tau_\nu{}^A \, C_\mu{}^B \bigr) \, \eta_{AB}\,, \\[2pt]
    \CB_{\mu\nu} & \rightarrow \CB_{\mu\nu} + \bigl( \tau_\mu{}^A \, C_\nu{}^B - \tau_\nu{}^A \, C_\mu{}^B \bigr) \, \epsilon_{AB}\,,
\end{align}
and
\begin{align}
    \tau_\mu & \rightarrow C \, \tau
        &%
    \lambda & \rightarrow C^{-1} \Bigl[ \lambda - \p \tilde{Y}^\mu \bigl( C_\mu{}^0 - C_\mu{}^1 \bigr) \Bigr]\,, \label{eq:Slambda} \\[2pt]
    \overline{\tau}_\mu & \rightarrow C \, \overline{\tau}_\mu\,,
        &%
    \overline{\lambda} & \rightarrow C^{-1} \Bigl[ \overline{\lambda} - \overline{\p} {\tilde{Y}}^\mu \bigl( C_\mu{}^0 + C_\mu{}^1 \bigr) \Bigr]\,. \label{eq:Slambdabar}
\end{align}
The parameters $C$ and $C_\mu{}^A$ have been introduced in \eqref{eq:deflambda}.\footnote{There is a derivative constraint that $C$ must satisfy: $E^\mu{}_{\!A'} \p_\mu C = 0$\,, where $E^\mu{}_{A'}$ is the inverse transverse Vielbein field in string Newton-Cartan geometry \cite{Bergshoeff:2019pij}.}
In presence of a dilaton field $\Phi$\,, we also have
\begin{align}
   	\Phi \rightarrow \Phi + \ln |C|\,.
\end{align}
\end{subequations}
In the broken phase, the embedding function $f^\mu$ develops a vacuum expectation value $f^\mu = f^\mu_0$ with $f^y_0 = \tilde{y}_0$ and $f^i_0 = X^i$, which breaks the string Newton-Cartan symmetry to the Bargmann symmetry, and also halves the Stueckelberg symmetry in \eqref{eq:unbrokensymmetries4}. 

Finally, we consider the same T-duality transformation of the worldvolume action of the spacetime-filling D$(d-1)$-brane in relativistic open string theory described by the action \eqref{eq:relaction} with its boundary conditions \eqref{eq:firstbdrycond}. For a single D-brane in spacetime, conformal invariance of the two-dimensional relativistic open string worldsheet leads to a nonlinear theory for the $U(1)$ connection living on the D-brane known as the Dirac-Born-Infeld (DBI) action. 
We start with the relativistic DBI action on a spacetime filling brane,
\be \label{eq:SDBI}
    S_\text{rDBI} = T_{d-1} \int d^{d} X^\mu \, e^{-\phi} \sqrt{-\det \CM_{\mu\nu}}\,,  
        \qquad
    \CM_{\mu\nu} \equiv g_{\mu\nu} + b_{\mu\nu} + F_{\mu\nu}\,.
\ee
Here, $X^\mu = (y\,, X^i)$\,, where $y$ is a lightlike isometry direction, with $g_{yy} = 0$\,.
We show that this action is T-dual to the DBI action on the D($d-$2)-brane in nonrelativistic open string theory \cite{Gomis:2020fui},\footnote{Also see \cite{Kluson:2019avy, Kluson:2020aoq} for a related worldvolume action and discussions on its T-duality transformation.}
\be \label{eq:tildeSDBI}
    \tilde{S}_\text{nrDBI} = \tilde{T}_{d-2} \int d^{d-1} X^i \, \tilde{\CL}_\text{nrDBI}\,, 
\ee
with
\be
    \tilde{\CL}_\text{nrDBI} = e^{-\Phi} \sqrt{- \det \!
    \begin{pmatrix}
        0 & \quad \p_j f^\mu \, \tau_\mu \\[2pt]
        \p_i f^\mu \, \overline{\tau}_\mu & \quad \p_i f^\mu \p_j f^\nu \bigl( H_{\mu\nu} + B_{\mu\nu} \bigr) + F_{ij}
    \end{pmatrix}}\,.
\ee
To show the equivalence of the two actions, we apply the Buscher rules \eqref{eq:HBAaction} and \eqref{eq:dilaton} to \eqref{eq:tildeSDBI} along with the explicit form of the embedding function,
\be
    f^y = \tilde{y}_0 - A_y\,,
        \qquad
    f^i = X^i,
\ee
which gives
\begin{align}
    \tilde{\CL}_\text{nrDBI} 
    & = e^{-\phi} \sqrt{- \det 
    \begin{pmatrix}
        0 & \quad \CM_{yj} \\[2pt]
        \CM_{iy} & \quad  \CM_{ij} + \lr \overline{C}_i + \overline{C}_y F_{yi} \rr \CM_{yj} + \CM_{iy} \lr C_j + C_y F_{yj} \rr
    \end{pmatrix}} \notag \\[2pt]
    & = e^{-\phi} \sqrt{- \det \CM_{\mu\nu}}\,.
\end{align}
This shows that the Lagrangian densities in $S_\text{rDBI}$ and $\tilde{S}_\text{nrDBI}$ are equal to each other. If the isometry $y$-direction is compactified on a circle of radius $R$\,, then, by identifying $2 \pi R \, T_{d-1} = \tilde{T}_{d-2}$\,, we find that $S_\text{rDBI} = \tilde{S}_\text{nrDBI}$\,.

\subsection{From Noncommutative to Nonrelativistic Open Strings} \label{sec:NCOStoNROS}

In this subsection, we discuss the generalization of the T-duality transformation that relates noncommutative and nonrelativistic open string theory in \S\ref{sec:nonrelnoncomm} to general background fields.
We start with the NCOS sigma model that generalizes the free action \eqref{eq:dualNCOS} in flat spacetime to a string Newton-Cartan geometry, $B$-field and spacetime-filling brane background, 
\begin{align}
    S_\text{NCOS} & = \frac{1}{4\pi\alpha'} \int_\Sigma d^2 \sigma \Bigl( \p_\alpha X^\mu \, \p^\alpha X^\nu h_{\mu\nu} - i \, \epsilon^{\alpha\beta} \p_\alpha X^\mu \, \p_\beta X^\nu b_{\mu\nu} + \lambda \, \overline{\p} X^\mu \, t_\mu + \overline{\lambda} \, \p X^\mu \,  \overline{t}_\mu \Bigr) \notag \\[2pt]
    & \quad + \frac{i}{2\pi\alpha'} \int_{\p\Sigma} d\tau \, A_\mu \, \p_\tau X^\mu\,.
\end{align}
The Neumann boundary conditions are 
\be \label{eq:Neumannbdry}
    h_{\mu\nu} \p_\sigma X^\nu + i \, \bigl( b_{\mu\nu} + F_{\mu\nu} \bigr) \p_\tau X^\nu + \tfrac{1}{2} \bigl( \lambda \, t_\mu + \overline{\lambda} \, \overline{t}_\mu \bigr) = 0\,.
\ee
In the neutral case, we write
\begin{align} \label{eq:NCOScurved}
    S_\text{NCOS} & = \frac{1}{4\pi\alpha'} \int_\Sigma d^2 \sigma \Bigl( \p_\alpha X^\mu \, \p^\alpha X^\nu h_{\mu\nu} - i \, \epsilon^{\alpha\beta} \p_\alpha X^\mu \, \p_\beta X^\nu \CB_{\mu\nu} \Bigr) \notag \\[2pt]
    & \quad + \frac{1}{4\pi\alpha'} \int_{\Sigma} d^2\sigma \, \Bigl( \lambda \, \overline{\p} X^\mu \, t_\mu + \overline{\lambda} \, \p X^\mu \,  \overline{t}_\mu \Bigr)\,,
\end{align}
where $h_{\mu\nu} = E_\mu{}^{A'} E_\nu{}^{A'} + \bigl( t_\mu{}^A m_\nu{}^B + t_\nu{}^A m_\mu{}^B \bigr) \eta_{AB}$ and $\CB_{\mu\nu} = b_{\mu\nu} + F_{\mu\nu}$\,. Here, $t_\mu{}^A$ is the longitudinal Vielbein field, $E_\mu{}^{A'}$ is the transverse Vielbein field, and $m_\mu{}^A$ is the gauge field associated with a noncentral extension in string Newton-Cartan algebra. See \cite{Bergshoeff:2018yvt}.

Consider a longitudinal lightlike Killing vector $\ell^\mu$ with 
\be \label{eq:longiso}
    \ell^\mu \, t_\mu \neq 0\,,
        \qquad
    \ell^\mu \, \overline{t}_\mu = 0\,,
        \qquad
    \ell^\mu E_\mu{}^{A'} = 0\,.
\ee
Define the adapted coordinates $X^\mu = (y, X^i)$ with respect to $\ell^\mu$, with $\ell^\mu \p_\mu = \p_y$\,. Gauging the isometry by introducing a flat gauge field $v_\alpha$\,, we find the gauged action
\begin{align} \label{eq:SNCOS3}
    S_\text{gauged} & = \frac{1}{4\pi\alpha'} \int_\Sigma d^2 \sigma \Bigl( D_\alpha X^\mu \, D^\alpha X^\nu h_{\mu\nu} \! - i \, \epsilon^{\alpha\beta} D_\alpha X^\mu \, D_\beta X^\nu \, \mathcal{B}_{\mu\nu} \! + \lambda \, \overline{D} X^\mu \, t_\mu \! + \overline{\lambda} \, \p X^i \,  \overline{t}_i \Bigr) \notag \\[2pt]
    & \quad - \frac{i}{2\pi\alpha'} \int_{\Sigma} d^2\sigma \, \epsilon^{\alpha\beta} \, \tilde{y} \, \p_\alpha v_\beta - \frac{i}{2\pi\alpha'} \int_{\p\Sigma} d\tau \, \tilde{y} \, v_\tau \,,
\end{align}
together with the Dirichlet boundary condition $\tilde{y} \, \big|_{\sigma = 0} = \tilde{y}_0$ for a constant $\tilde{y}_0$\,, which is required by the gauge invariance.
Here, $D_\alpha X^\mu = \p_\alpha X^\mu + k^\mu \, v_\alpha$\,.
The gauged version of the Neumann boundary conditions in \eqref{eq:Neumannbdry} is
\be \label{eq:Neumann3gauge}
   h_{\mu\nu} D_\sigma X^\nu + i \, \CB_{\mu\nu} \, D_\tau X^\nu + \tfrac{1}{2} \bigl( \lambda \, t_\mu + \overline{\lambda} \, \overline{t}_\mu \bigr) = 0\,.
\ee

Now, we perform a T-duality transformation along the isometry $y$-direction by integrating out $v_\alpha$ in the action $S_\text{NCOS}$ from \eqref{eq:SNCOS3}. This can be done by first writing down the equation of motion for $v_\alpha$ as
\begin{subequations} \label{eq:vtauvsigma}
\begin{align}
    v_\tau & = - \frac{1}{h_{yy}} \Bigl[  \, h_{y\nu} \, \p_\tau X^\nu - i \, \bigl( \mathcal{B}_{y\nu} \, \p_\sigma X^\nu + \p_\sigma \tilde y \bigr) + \tfrac{i}{2} \, \lambda \, t_y \Bigr]\,, \\[2pt]
    v_\sigma & = - \frac{1}{h_{yy}} \Bigl[ \, h_{y\nu} \, \p_\sigma X^\nu + i \,\bigl( \mathcal{B}_{y\nu} \, \p_\tau X^\nu + \p_\tau \tilde y \bigr) + \tfrac{1}{2} \, \lambda \, t_y \Bigr]\,.
\end{align}
\end{subequations}
Plugging $v_\alpha$ into \eqref{eq:SNCOS3}, and taking the redefinition $\tilde{X}^\mu = \bigl( \tilde{y}\,, X^i \bigr)$ and
\be
    \lambda = - h_{yy} \, t^{-1}_y \, C \, \tilde{\lambda} - {D} \tilde{X}^\mu \, \overline{C}_\mu\,,
        \qquad
	D_\alpha \tilde{X}^\mu = \p_\alpha \tilde{X}^\mu - \ell^\mu \, \p_\alpha X^i \, \p_i A_y\,,
\ee
with $C \neq 0$\,, 
we find the dual action
\begin{align} \label{eq:dualNROS3}
    \tilde{S}_\text{NROS} & = \frac{1}{4\pi\alpha'} \int_\Sigma d^2 \sigma \Bigl( D_\alpha \tilde{X}^\mu \, D^\alpha \tilde{X}^\nu \, H_{\mu\nu} - i \, \epsilon^{\alpha\beta} D_\alpha \tilde{X}^\mu \, D_\beta \tilde{X}^\nu \, B_{\mu\nu} \Bigr) \notag \\[2pt]
    & \quad + \frac{1}{4\pi\alpha'} \int_\Sigma d^2 \sigma \Bigl( \tilde{\lambda} \, \overline{D} \tilde{X}^\mu \, \tau_\mu + \overline{\lambda} \, \p X^i \,  \overline{\tau}_i \Bigr) \notag \\[2pt]
    & \quad + \frac{i}{2\pi\alpha'} \int_{\p\Sigma} d\tau \lr A_i \, \p_\tau X^i + \tilde{y}_0 \, \p_\tau y \rr,
\end{align}
where
\begin{subequations} \label{eq:lightlikeBusher}
\begin{align}
    H_{yi} & = \frac{b_{yi}}{h_{yy}} + \frac{t_y \lr \tau_y \, \overline{C}_i + \tau_i \, \overline{C}_{y}  \rr}{2 \, h_{yy}  \, C}\,,  
        \qquad\,\,\,\,\,\, %
    \tau_y = C\,, 
        \qquad\,\,\,\,\,\,
    \overline{\tau}_i = \overline{t}_i\,, \\[2pt]
    B_{yi} & = \frac{h_{yi}}{h_{yy}} - \frac{t_y \lr \tau_y \, \overline{C}_i - \tau_i \, \overline{C}_y \rr}{2 \, h_{yy} \, C}\,, 
        \qquad\,\,\,\,\,\, %
    \tau_i = C \, \frac{\bigl( h_{yi} + b_{yi} \bigr) \, t_y - h_{yy} \, t_i}{t_y}\,, \\[2pt] 
    H_{ij} & = h_{ij} + \frac{b_{yi} \, b_{yj} - h_{yi} \, h_{yj}}{h_{yy}} + \frac{t_y \lr \tau_i \, \overline{C}_j + \tau_j \, \overline{C}_{i} \rr}{2 \, h_{yy} \, C} \,, 
        \qquad\!\! %
    H_{yy} = \frac{1 + t_y \, \overline{C}_y}{h_{yy}}\,, \\[2pt]
    B_{ij} & = b_{ij} + \frac{b_{yi} \, h_{yj} - b_{yj} \, h_{yi}}{h_{yy}} - \frac{t_y \lr \tau_i \, \overline{C}_j - \tau_j \, \overline{C}_i \rr}{2 \, h_{yy} \, C}\,.
\end{align}
When the dilaton $\phi$ is included, the dual dilaton $\Phi$ is
\be
    \Phi = \phi + \tfrac{1}{2} \log \bigl( t_y^{-1} C \bigr)\,.
\ee
\end{subequations}
If one takes
\be
    C = - \overline{C}_y = \frac{1}{t_y}\,,
        \qquad
    \overline{C}_i =  \frac{\bigl(h_{yi} - b_{yi} \bigr) \, t_y - h_{yy} \, t_i}{t_{y}^2}\,,
\ee
then \eqref{eq:lightlikeBusher} becomes
\begin{subequations}
\begin{align}
    H_{ij} & = h_{ij} + \frac{h_{yy} \, t_i \, t_j - \bigl( h_{yi} \, t_j + h_{yj} \, t_i \bigr) \, t_y}{t_y^2}\,, 
        \qquad
    \Phi = \phi - \log |t_y|\,, \\[2pt]
    H_{y\mu} & = 0\,,  
        \qquad\,\,\,\,\,\, %
    B_{yi} = \frac{t_i}{t_y}\,, 
        \qquad\,\,\,\,\,\, %
    B_{ij} = b_{ij} + \frac{b_{yi} \, t_j - b_{yj} \, t_i}{t_{y}}\,, \\[2pt]
    \tau_y & = \frac{1}{t_y}\,, 
        \qquad\,\,\,\,\,\,\, %
    \overline{\tau}_i = \overline{t}_i\,,
        \qquad\,\,\,\,\,\,\,\,\,\,\,
    \tau_i = \frac{b_{yi} \, t_y - h_{yy} \, t_i + h_{yi} \, t_y}{t_y^2}\,,
\end{align}
\end{subequations}
which recovers the Buscher rules in \cite{Bergshoeff:2018yvt}. Note that this set of Buscher rules is non-singular in the limit $h_{yy} \rightarrow 0$\,. Plugging \eqref{eq:dualNROS3} into the boundary condition \eqref{eq:Neumann3gauge}, we find the following dual boundary conditions:
\begin{subequations} \label{eq:dualbdry3}
\begin{align} 
    \p_\tau \tilde{y} & = 0\,, \\[2pt]
    D_i \tilde{X}^\mu \ls H_{\mu\nu} \, D_\sigma \tilde{X}^\nu + i \, B_{\mu\nu} \, D_\tau \tilde{X}^\nu + \tfrac{1}{2} \lr \tilde{\lambda} \, \tau_\mu + \overline{\lambda} \, \overline{\tau}_\mu \rr \rs + i \, F_{ij} \, \p_\tau X^j  & = 0\,,
\end{align}
\end{subequations}
where $D_i \tilde{X}^\mu = \delta^\mu_i - \ell^\mu \, \p_i A_y$\,. We also used the constraint $\overline{D} \tilde{X}^\mu \, \tau_\mu = 0$ imposed by the Lagrange multiplier $\tilde{\lambda}$ in \eqref{eq:dualNROS3}.
The dual theory is nonrelativistic open string theory. The dual $\tilde{y}$-direction is a lightlike isometry direction, in the sense that the Vielbeine fields satisfy
\be \label{eq:llc}
    \ell^\mu \tau_\mu \neq 0\,,
        \qquad
    \ell^\mu \, \overline{\tau}_\mu = 0\,,
        \qquad
    \ell^\mu E_\mu{}^{A'} = 0\,.
\ee
In this dual theory, there is a D($d-$2)-brane transverse to the $\tilde{y}$-direction. One may also absorb the Nambu-Goldstone boson $A_y$ into the definition of a collective coordinate $\tilde{Y} = \tilde{y}_0 - A_y$ and the discussion goes the same as in \S\ref{sec:RtoNR}.

Finally, we perform the same T-duality transformation on the DBI type action for the spacetime-filling D$(d-1)$-brane in NCOS,
\be \label{eq:ncDBI}
    S_\text{ncDBI} = T_{d-1} \int d^d X^\mu \, \CL_\text{ncDBI}\,,
\ee
with
\be
    \CL_\text{ncDBI} = e^{-\phi} \sqrt{\det \begin{pmatrix}
        0 & \quad t_\mu \\[2pt]
        \overline{t}_\mu & \quad h_{\mu\nu} + b_{\mu\nu} + F_{\mu\nu}
    \end{pmatrix}}\,.
\ee
In presence of the lightlike isometry direction $y$\,, we rewrite $S_\text{ncDBI}$ as
\begin{align} \label{eq:LncDBItc}
    \CL_\text{ncDBI} & = e^{-\phi} \sqrt{\det \begin{pmatrix}
        0 & \quad t_y & \quad t_j \\[2pt]
        0 & \quad \CM_{yy} & \quad \CM_{yj} \\[2pt]
        \overline{t}_i & \CM_{iy} & \quad \CM_{ij}
    \end{pmatrix}}
    = e^{-\phi} \sqrt{- \det \begin{pmatrix}
        t_y & \quad 0 & \quad t_j \\[2pt]
        \CM_{yy} & \quad 0 & \quad \CM_{yj} \\[2pt]
        \CM_{iy} & \quad \overline{t}_i & \quad \CM_{ij}
    \end{pmatrix}} \notag \\[2pt]
    & = e^{-\phi} \sqrt{- \det 
    \begin{pmatrix}
        0 & \quad \CM_{yj} \, t_y - \CM_{yy} \, t_j \\[2pt]
        \overline{t}_i & \quad \CM_{ij} \, t_y - \CM_{iy} \, t_j
    \end{pmatrix}}\,,
\end{align}
where we defined $\CM_{\mu\nu} = h_{\mu\nu} + b_{\mu\nu} + F_{\mu\nu}$\,. 
On the other hand, the worldvolume action for a D-brane transverse to the longitudinal lightlike isometry direction in NROS is
\be \label{eq:tildeSnrDBI}
    \tilde{S}_\text{nrDBI} = \tilde{T}_{d-2} \int d^{d-1} X^i \, \tilde{\CL}_\text{nrDBI}\,, 
\ee
with
\be
    \tilde{\CL}_\text{nrDBI} = e^{-\Phi} \sqrt{- \det \!
    \begin{pmatrix}
        0 & \quad \p_j f^\mu \, \tau_\mu \\[2pt]
        \p_i f^\mu \, \overline{\tau}_\mu & \quad \p_i f^\mu \p_j f^\nu \bigl( H_{\mu\nu} + B_{\mu\nu} \bigr) + F_{ij}
    \end{pmatrix}}\,,
\ee
where $f^y = \tilde{y}_0 - A_y$ and $f^i = X^i$.
Plugging the Buscher rules \eqref{eq:lightlikeBusher} into \eqref{eq:tildeSnrDBI} and comparing with \eqref{eq:LncDBItc}, we find $\tilde{\CL}_\text{nrDBI} = \CL_\text{ncDBI}$\,.
This proves that the two worldvolume actions, \eqref{eq:ncDBI} and \eqref{eq:tildeSnrDBI}, are T-dual to each other.

\subsection{From Noncommutative to Relativistic Open Strings} \label{eq:nctorc}

Finally, we consider the generalization to arbitrary background fields of the T-duality transformation in \S\ref{sec:noncommtorel} that relates noncommutative to relativistic open string theory. We start with noncommutative open string theory in string Newton-Cartan geometry with a lightlike isometry in the longitudinal sector, which is described by the sigma model in \eqref{eq:NCOScurved} together with the boundary condition in \eqref{eq:Neumannbdry}.
Instead of taking the longitudinal lightlike Killing vector in \eqref{eq:longiso}, we now assume that there exists a longitudinal spatial Killing vector $k^\mu$\,, with
\be
    k^\mu \, t_\mu{}^0 = 0\,,
        \qquad
    k^\mu \, t_\mu{}^1 \neq 0\,,
        \qquad
    k^\mu E_\mu{}^{A'} = 0\,.
\ee
Gauging the isometry in \eqref{eq:NCOScurved} by introducing a flat gauge field $v_\alpha$\,, we find
\begin{align} \label{eq:SNCOStorel}
    S_\text{NCOS} & = \frac{1}{4\pi\alpha'} \int_\Sigma d^2 \sigma \Bigl( D_\alpha X^\mu \, D^\alpha X^\nu h_{\mu\nu} - i \, \epsilon^{\alpha\beta} D_\alpha X^\mu \, D_\beta X^\nu \, \CB_{\mu\nu} \Bigr) \notag \\[2pt]
    & \quad + \frac{1}{4\pi\alpha'} \int_\Sigma d^2 \sigma \Bigl( 
    \lambda \, \overline{D} X^\mu \, t_\mu + \overline{\lambda} \, D X^\mu \, \overline{t}_\mu  \Bigr) \notag \\[2pt]
    & \quad - \frac{i}{2\pi\alpha'} \int_{\Sigma} d^2\sigma \, \epsilon^{\alpha\beta} \, \tilde{y} \, \p_\alpha v_\beta - \frac{i}{2\pi\alpha'} \int_{\p\Sigma} d\tau \, \tilde{y} \, v_\tau \,,
\end{align}
where $D_\alpha X^\mu = \p_\alpha X^\mu + k^\mu \, v_\alpha$ and $\tilde{y}$ satisfies the Dirichlet boundary condition $\tilde{y} \, \big|_{\sigma = 0} = \tilde{y}_0$ for a constant $\tilde{y}_0$\,, required by the gauge invariance.
The gauged version of the boundary condition \eqref{eq:Neumannbdry} is
\be \label{eq:NeumannNCOSgauged}
     h_{\mu\nu} \, D_\sigma X^\nu + i \, \bigl( b_{\mu\nu} + F_{\mu\nu} \bigr) D_\tau X^\nu + \tfrac{1}{2} \bigl( \lambda \, t_\mu + \overline{\lambda} \, \overline{t}_\mu \bigr) = 0\,.
\ee
Varying with respect to $v_\alpha$ in \eqref{eq:SNCOStorel}, we find
\begin{subequations} \label{eq:lambdarules2}
\begin{align} 
    \lambda & = - \frac{1}{t_y} \bigl( - \p \, \tilde{y} - \mathcal{B}_{yi} \, \p X^i + h_{y\mu} \, D X^\mu \bigr)\,, \\[2pt]
    \overline{\lambda} & = - \frac{1}{\overline{t}_y} \bigl( \, \overline{\p} \, \tilde{y} + \mathcal{B}_{yi} \, \overline{\p} X^i + h_{y\mu} \, \overline{D} X^\mu \bigr)\,.
\end{align}
\end{subequations}
where $D = D_\sigma - i D_\tau$ and $\overline{D} = D_\sigma + i D_\tau$\,. Also note that varying \eqref{eq:SNCOStorel} with respect to $\lambda$ and $\overline{\lambda}$ gives rise to the equations of motion for $v_\alpha$\,, with
\be \label{eq:vrules2}
    v_\alpha = - \p_\alpha y - \frac{1}{t_y{}^1} \, \bigl( \p_\alpha X^i \, t_i{}^1 - i \, \epsilon_\alpha{}^\beta \, \p_\beta X^i \, t_i{}^0 \bigr)\,.
\ee

We perform a T-duality transformation along the isometry $y$ direction, with $\p_y = k^\mu \p_\mu$\,, by integrating out the auxiliary gauge field $v_\alpha$\,. This amounts to using \eqref{eq:lambdarules2} and \eqref{eq:vrules2} to eliminate $\lambda$\,, $\overline{\lambda}$ and $v_\alpha$ in \eqref{eq:SNCOStorel}, 
which leads to the dual action
\begin{align} \label{eq:dualROS2}
    \tilde{S}_\text{ROS} & = \frac{1}{4\pi\alpha'} \int_\Sigma d^2 \sigma \Bigl( D_\alpha {\tilde{X}}^{\mu} \, D^\alpha {\tilde{X}}^{\nu} \, G_{\mu\nu} - i \, \epsilon^{\alpha\beta} \, D_\alpha {\tilde{X}}^\mu \, D_\beta {\tilde{X}}^\nu \, B_{\mu\nu} \Bigr) 
 \notag \\[2pt]
    & \quad + \frac{i}{2\pi\alpha'} \int_{\p\Sigma} d\tau \, \Bigl( A_i \, \p_\tau X^i - \tilde{y} \, \p_\tau y \Bigr),
\end{align}
with the Buscher rules \cite{Bergshoeff:2018yvt}
\begin{subequations} \label{eq:ROSbdryconds}
\begin{align}
    G_{yy} & = 0\,, \\[2pt]
    G_{yi} & = \frac{t_i{}^A \, t_y{}^B \epsilon_{AB}}{t_{yy}}\,, 
        \hspace{3cm}
    B_{yi} = \frac{t_{yi}}{t_{yy}}\,, \\[2pt]
    G_{ij} & = h_{ij} + \frac{\bigl( b_{yi} \, t_j{}^A + b_{yj} \, t_i{}^A \bigr) \, t_y{}^B \epsilon_{AB} + h_{yy} t_{ij} - h_{yi} \, t_{yj} - h_{yj} \, t_{yi}}{t_{yy}}\,, \\[2pt]
    B_{ij} & = b_{ij} + \frac{b_{yi} \, t_{yj} - b_{yj} \, t_{yi} - \bigl( h_{yy} \, t_i{}^A \, t_j{}^B - h_{yi} \, t_y{}^A \, t_j{}^B + h_{yj} \, t_y{}^A \, t_i{}^B \bigr) \, \epsilon_{AB}}{t_{yy}}\,,
\end{align}
\end{subequations}
where $t_{\mu\nu} = t_\mu{}^A \, t_\nu{}^B \, \eta_{AB}$\,. In addition, the dilaton field $\phi$\,, when included, transforms into $\Phi$ with
\be \label{eq:PhiBuscher2}
    \Phi = \phi - \tfrac{1}{2} \ln t_{yy}\,.
\ee
Plugging \eqref{eq:lambdarules2} and \eqref{eq:vrules2} into the boundary conditions in \eqref{eq:Neumannbdry}, we find the following dual boundary conditions:
\begin{align}\label{ROSBC}
    \p_\tau \tilde{y} = 0\,, 
        \qquad %
    D_i \tilde{X}^\mu \lr G_{\mu\nu} \, D_\sigma \tilde{X}^\nu + i \, B_{\mu\nu} \, D_\tau \tilde{X}^\nu  \rr + i \, F_{ij} \, \p_\tau X^j = 0\,. 
\end{align}
The dual theory is described by the sigma model for relativistic strings, whose target space is Riemannian. The dual $\tilde{y}$-direction is a lightlike isometry direction due to the condition $G_{yy} = 0$\,. In this dual theory, there is a D($d-2$)-brane that is transverse to the compactified lightlike $\tilde{y}$-direction, which implies that the dual theory is the DLCQ of relativistic open string theory.

Introducing the collective coordinate $\tilde Y = \tilde{y} - A_y$\,,
we rewrite \eqref{eq:dualROS2} as
\begin{align}
    \tilde{S}_\text{ROS} & = \frac{1}{4\pi\alpha'} \int_\Sigma d^2 \sigma \Bigl( \p_\alpha \tilde {Y}^{\mu} \, \p^\alpha \tilde {Y}^{\nu} \, G_{\mu\nu} - i \, \epsilon^{\alpha\beta} \, \p_\alpha \tilde {Y}^\mu \, \p_\beta \tilde {Y}^\nu \, B_{\mu\nu} \Bigr) \notag \\[2pt]
    & \quad + \frac{i}{2\pi\alpha'} \int_{\p\Sigma} d\tau  \, \Bigl( A_i \, \p_\tau X^i - \tilde{y}_0 \, \p_\tau y \Bigr),
\end{align}
where $\tilde Y^\mu = (\tilde Y, X^i)$\,. The boundary conditions in \eqref{ROSBC} become
\be \label{DualROS}
    \p_\tau \tilde Y + \p_\tau A_y = 0\,,
        \qquad
    \p_i \tilde Y^\mu \, G_{\mu\nu} \, \p_\sigma \tilde Y^\nu + i \, \bigl( \p_i \tilde Y^\mu B_{\mu\nu} \, \p_j \tilde Y^\nu + F_{ij} \bigr) \, \p_\tau X^j = 0\,. 
\ee
Restricting the support of $A_\mu$ to be on the boundary requires $\tilde Y = \tilde{y}$ in the bulk and $\tilde Y \big|_{\sigma = 0} = \tilde{y}^{}_0 - A_y$ on the boundary.
In a covariant form, the boundary conditions are
\begin{subequations}
\begin{align}
    & \tilde Y^\mu = f^\mu(X^i)\,,
        \qquad
    \p_\sigma \tilde Y^\mu \, \tau_\mu{}^A = i \, \epsilon^A{}_B \, \p_\tau X^i \, \tau_i{}^B\,, \\[2pt]
    & \p_i f^\mu G_{\mu\nu} \, \p_\sigma \tilde Y^\nu + i \, \CF_{ij} \, \p_\tau X^j = 0\,,
\end{align}
\end{subequations}
where $f^\mu (X^i)$ is the embedding function of the D-brane and $\CF_{ij} = \p_i f^\mu \, B_{\mu\nu} \, \p_j f^\nu + F_{ij}$\,.

Finally, we consider the T-duality transformation acting on the worldvolme actions for D-branes in spacetime. We start with the DBI action for the D($d-2$)-brane in relativistic open string theory is
\begin{align} \label{eq:rDBI2}
    \tilde{S}_\text{rDBI} = \tilde{T}_{d-2} \int d^{d-1} X^i \, e^{-\Phi} \sqrt{-\det \ls \p_i f^\mu \bigl( G_{\mu\nu} + B_{\mu\nu} \bigr) \p_j f^\nu + F_{ij} \rs}.
\end{align}
where $f^y = \tilde{y}_0 - A_y$ and $f^i = X^i$. Plugging the Buscher rules \eqref{eq:ROSbdryconds} and \eqref{eq:PhiBuscher2} into \eqref{eq:rDBI2}, and using $f^\mu = X^\mu - k^\mu A_y$\,, we find
\be
    S_\text{rDBI} = T_{d-2} \int d^{d-1} X \, e^{-{\phi}} \sqrt{- \det \CN_{ij}}\,, 
\ee
with
\be
    \CN_{ij} = - \overline{t}_y \, t_y \, \CM_{ij} + {\overline{t}_i \, t_y} \, \CM_{yj} + \CM_{iy} \, {t_j \, \overline{t}_y} - \CM_{yy} \, {\overline{t}_i \, t_j}\,,
\ee
where $\CM_{\mu\nu} = h_{\mu\nu} + b_{\mu\nu} + F_{\mu\nu}$\,.
Note that
\begin{align}
    & \quad - \det
    \begin{pmatrix}
        0 & \,\, t_y & \,\, t_j \\[2pt]
        \overline{t}_y & \,\, \CM_{yy} & \,\, \CM_{yj} \\[2pt]
        \overline{t}_i & \,\, \CM_{iy} & \,\, \CM_{ij}
    \end{pmatrix}
        =
     \det
    \begin{pmatrix}
        t_y & \,\, 0 & \,\, t_j \\[2pt]
        \CM_{yy} & \,\, \overline{t}_y & \,\, \CM_{yj} \\[2pt]
        \CM_{iy} & \,\, \overline{t}_i & \,\, \CM_{ij}
    \end{pmatrix} \notag \\[2pt]
    & =  \det
    \begin{pmatrix}
        \overline{t}_y & \,\, \CM_{yj} \, t_y - \CM_{yy} \, t_j \\[2pt]
        \overline{t}_i & \,\, \CM_{ij} \, t_y - \CM_{iy} \, t_j
    \end{pmatrix}
    = - \det \CN_{ij}\,.
\end{align}
Therefore, we find that the DBI type action for the spacetime-filling brane in NCOS is
\begin{align} \label{eq:SncDBI}
    S_\text{ncDBI} = T_{d-1} \int d^d X^\mu \, e^{-\phi} \sqrt{- \det \begin{pmatrix}
        0 & \quad t_\mu \\[2pt]
        \overline{t}_\mu & \quad h_{\mu\nu} + b_{\mu\nu} + F_{\mu\nu}
    \end{pmatrix}}\,.
\end{align}
We conclude that the worldvolume actions $\tilde{S}_\text{rDBI}$ in \eqref{eq:rDBI2} and $S_\text{ncDBI}$ in \eqref{eq:SncDBI} are T-dual to each other.

\acknowledgments

This research is supported in part by Perimeter Institute for Theoretical Physics. Research at Perimeter Institute is supported in part by the Government of Canada through the Department of Innovation,
Science and Economic Development Canada and by the Province of Ontario through the Ministry of Colleges and Universities.

\newpage

\bibliographystyle{JHEP}
\bibliography{nrostd}

\end{document}